\documentclass[aps,pre,twocolumn,superscriptaddress,floatfix]{revtex4}

\usepackage{amsthm,amssymb,url,amsmath}

\usepackage{hyperref}
\usepackage{graphicx}
\usepackage{color}
\usepackage{multirow}
\usepackage{tabularx}
\usepackage{textcomp}
\usepackage{relsize}
\usepackage{csquotes}
\usepackage{setspace}
\usepackage{enumerate}

\newcommand{\norm}[1]{\left|\left| #1 \right|\right|}

\newcommand{\multititle}[1]{\begin{minipage}{0.15\textwidth}\begin{flushleft}
      \smaller \textbf{#1}  \end{flushleft}\end{minipage}}

\begin{document}


\title{The `who' and `what' of \#diabetes on Twitter}

\author{Mariano Beguerisse-D\'iaz} \email{beguerisse@maths.ox.ac.uk}
\affiliation{Department of Mathematics, Imperial College London, UK}
\affiliation{Mathematical Institute, University of Oxford, UK}

\author{Amy K. McLennan}
\affiliation{School of Anthropology and Museum
  Ethnography, University of Oxford, UK}

\author{Guillermo Gardu\~no-Hern\'andez}
\affiliation{Sinnia, Mexico City, Mexico}

\author{Mauricio Barahona} \email{m.barahona@imperial.ac.uk}
\affiliation{Department of Mathematics, Imperial College London, UK}

\author{Stanley J. Ulijaszek} \email{stanley.ulijaszek@anthro.ox.ac.uk}
\affiliation{School of
  Anthropology and Museum Ethnography, University of Oxford, UK}

\begin{abstract}
  Social media are being increasingly used for health promotion, yet the landscape of users, messages and interactions in such fora is poorly understood. Studies of social media and diabetes have focused mostly on patients, or public agencies addressing it, but have not looked broadly at all the participants or the diversity of content they contribute. We study Twitter conversations about diabetes through the systematic analysis of 2.5 million tweets collected over 8 months and the interactions between their authors. We address three questions: (1) what themes arise in these tweets?; (2) who are the most influential users?; (3) which type of users contribute to which themes? We answer these questions using a mixed-methods approach, integrating techniques from anthropology, network science and information retrieval such as thematic coding, temporal network analysis, and community and topic detection.  Diabetes-related tweets fall within broad thematic groups: health information, news, social interaction, and commercial. At the same time, humorous messages and references to popular culture appear consistently, more than any other type of tweet. We classify authors according to their temporal `hub' and `authority' scores.  Whereas the hub landscape is diffuse and fluid over time, top authorities are highly persistent across time and comprise bloggers, advocacy groups and NGOs related to diabetes, as well as for-profit entities without specific diabetes expertise. Top authorities fall into seven interest communities as derived from their Twitter follower network.  Our findings have implications for public health professionals and policy makers who seek to use social media as an engagement tool and to inform policy design.
\end{abstract}

\maketitle

\section{Introduction}

Of an estimated three billion internet users around the world (over
40\% of the global population), approximately 310 million actively use
the social networking site Twitter~\footnote{According to Twitter's own
  figures from July 2016}. Since Twitter's launch in 2006, the
platform and its users have been the protagonists of several major
events (e.g., the Arab Spring, protest movements in Iran, Spain and
elsewhere ~\cite{Beguerisse2014, Denef2013,
  Gonzalez-Bailon2011,Gonzalez-Bailon2016, Morales2012, Tonkin2012,
  Zhou2010}), and it has become a prominent venue for companies,
personalities, and ordinary people to broadcast news and events, send
public messages, express opinions, and socialise~\cite{Kwak2010,
  Wu2011}.

There is growing interest in the potential uses for Twitter, and other
social media, in public and population health. Research has been
carried out on the use of Twitter for epidemiological applications and
public health surveillance. For instance, influenza
spread~\cite{Culotta2010, Signorini2011}, contagious disease
outbreaks~\cite{Garcia-Herranz2014} and tobacco use~\cite{Prier2011}
have been mapped using public data from Twitter. Geographic or spatial
risk factors have also been elucidated using
Twitter~\cite{Paul2011,Llorente2015}.  This medium is being
increasingly used for health information sharing~\cite{Scanfeld2010},
primary care, delivery of health support, primary prevention, and
public health education~\cite{Hawn2009,Heaivilin2011,King2013,
  PHE2015, PHE2015a}.

Health promotion has typically drawn on unidirectional, top-down
social marketing and advertising strategies to disseminate
health-related messages to a wide audience~\cite{Lupton2012}. The
recent use of Twitter in health promotion coincides with the influx of
commercial agencies into social media to sell commodities, trends and
ideas~\cite{Lupton2012}.  The confluence of health promotion and
commercial agents on Twitter has led to a cacophony of short
health-related messages directed at users and passed through social
networks. Some users who create and disseminate content have the
exclusive aim of promoting health, while others may have additional or
alternative aims (e.g., commercial interest, self promotion). Yet
this complex environment is rarely acknowledged in public health
promotion research, reviews and planning. Instead, public health
marketing strategies and evaluations focus on the quantity of
messages disseminated by health authorities over time, the number of
followers an account has attracted, and the design of new
slogans~\cite{PHE2015,PHE2015a}. There is little critical analysis of
the wider landscape of, and relationships underpinning, health-related
content on Twitter, or of the impact that messages have in the broad
user base.

\begin{figure*}[t]
  \centerline{\includegraphics[width=\textwidth]
    {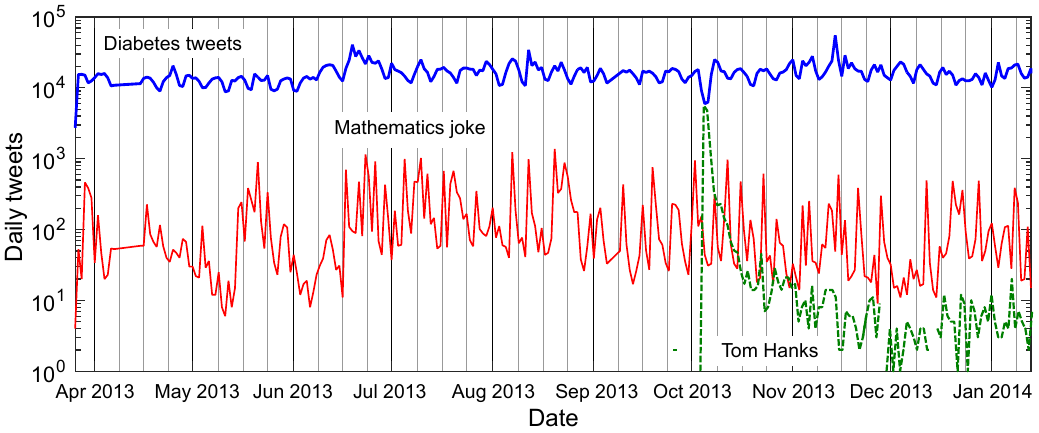}}
  \caption{The number of daily tweets in English containing the term
    `diabetes' (blue line), the number of daily tweets containing some
    version of the `Mathematics joke', a particularly recurrent tweet
    (red line, see also Table~\ref{tab:Recurrent}), and the number of
    tweets mentioning the actor Tom Hanks (green dashed-line), who
    revealed his diabetes in October 2013.}
  \label{fig:num_tweets}
\end{figure*}

In this paper, we investigate messages on Twitter (`what' is
discussed, and by `whom') relating to a single clinical and public
health concern: diabetes. Diabetes is a clinical condition associated
with blood sugar regulation by the hormone insulin and other endocrine
factors. Complications associated with unmanaged or poorly-managed
diabetes can typically affect the heart, blood vessels, eyes, kidneys
and nerves. There are two types of diabetes: Type 1 (often abbreviated
as T1 or T1D on Twitter) occurs when the pancreas does not produce
sufficient insulin; Type 2 (T2 or T2D) develops when the body does not
effectively respond to insulin. In 2012, the global prevalence of
diabetes of both types was estimated to be nine percent among adults
aged 18 years and over~\cite{WHO2015}. However, this is likely to be
an underestimate; in the United States alone, 27.8\% of all diabetes
cases are thought to be undiagnosed~\cite{CDC2014}. Globally, around
10\% of all people with diabetes have T1, which typically requires the
daily administration of insulin. The remaining 90\% have diagnoses of
T2, which is commonly associated with obesity, poor nutrition and
physical inactivity. Both types have genetic components, although no
known genes directly cause diabetes by themselves~\cite{WHO2015}. T2
is managed predominantly through health education, diet activity
change, and weight loss; treatments also may include medication, and
in more advanced cases, regular insulin
administration~\cite{WHO2015}. People who have diabetes are encouraged
to self-surveil their blood sugar levels and self-manage their
lifestyles and insulin levels (through diet, physical activity, and
medications, where prescribed) in order to control their blood sugar
levels and minimise related health complications such as those
described above.

Previous research on diabetes and social media has focused on the
possibilities that the internet and social media open up for both
self-management~\cite{Shaw2011,ElGayar2013} and clinical
management~\cite{Kaufman2010, Gough2015, Harris2013} for people living
with the disease.  Others have focused on the dissemination of
information relating to diabetes-specific outreach
events~\cite{Desai2012} or from specific platforms~\cite{Greene2011}.
Our approach is different; instead of assuming that patients,
clinicians and health promotion organisations are the main
protagonists in the Twitter conversation about diabetes, we place all
stakeholders on an initial equal footing with respect to Twitter
exchanges around the broad category `diabetes'.

We compiled over 2.5 million English-language \textit{tweets}
containing the term `diabetes', which were generated by more than one
million users over a period of 8 months (March 2013 to January
2014). These are the type of messages that users encounter when they
search for the term `diabetes' on Twitter, or click on the hashtag
{\it \#diabetes}.  We analysed these tweets (public messages
comprising of text strings with a maximum length of 140 characters),
using a mixed-methods approach that combines mathematical and
computational techniques with anthropological analysis. We used tools
from data and network science to detect patterns in social
interactions~\cite{Kleinberg1999, Newman2010, Beguerisse2014} and to
extract topics from the messages~\cite{Lancichinetti2015,
  Manning2008}.  We then interrogated these patterns using discourse
analysis approaches from anthropology~\cite{Wilson2002, Bowen2008,
  Braun2006}, which permit the elaboration of themes, personalities,
and contexts.  As a result, we extract and classify the topics that
appear in the messages, and identify the important participants in the
conversations (which include patients, practitioners, public health
authorities, commercial entities and others).

Researchers have called for improving our understanding of social media
to inform its use in public health policy making and
practice~\cite{Harris2013}. Our work contributes to this understanding
by addressing three questions. First, {\it what are the main themes of
  messages posted on Twitter that contain the term `diabetes'?} To
answer this question, we extract the topics that appear in the tweets
by processing word co-occurrence networks, and analyse their content,
participants, and evolution. Second, {\it who talks about diabetes on
  Twitter and in what capacity?} To answer this question, we
investigate the Twitter users who drive conversations by analysing
networks of interactions (posting, sharing/retweeting, following)
among the users.  And third, {\it which users contribute content to
  which topics and themes?}  We answer this question by examining the
type of accounts that post in the different themes.  We discuss the
relevance of this interdisciplinary research for public health
professionals and policy.  The methods employed in this project are
general and may be applied in other studies where similar data become
available.

\section{Methods}

\subsection{Data from Twitter}

We collected every tweet containing the term `diabetes' (2,698,114
tweets in total), posted between 26 March 2013 and 19 January 2014 by
1,202,143 different users (Fig.~\ref{fig:num_tweets}).  We also
collected information about `retweets'---a retweet event is when a
user re-broadcasts a message (or `status') originally posted by
another user and which is disseminated to his or her followers.  Each
retweet is a time-annotated interaction between two users: the target
(the author of the original tweet, or the target of the attention) and
the source (the user who retweeted the message, or the source of the
attention).  We recorded 1,219,282 retweets from June 2013 up until
when the data collection ended. We also recorded 41,582
friend-follower relationships among a select subset of the users along
with their {\it Twitter biographies}, messages of at most 140
character in which users can describe themselves.  All data were
collected by Sinnia using Twitter Gnip PowerTrack
API~\footnote{\url{https://gnip.com/realtime/powertrack/}}. We have
made available a list containing tweet IDs used in this study (see
Data Statement).

There is an ongoing debate about the ethical implications of using
Twitter data for research. Some authors maintain that the lack of
complicated privacy settings on Twitter means that messages placed in
the public domain are intended to be there; alternatively, other
authors consider that posting tweets should not be interpreted as
permission to use tweets for research~\cite{Zimmer2014}. We believe
that the topic, analysis and results presented here serve the public
interest and pose no risk to users.  None of the tweets we analyse and
reproduce here contain notable amounts of sensitive or private
material. Indeed, the most prominent users in our data set also
maintain other online profiles and produce tweets for public
consumption.

\subsection{Construction of the different networks}

The number of `diabetes' tweets follows weekly cycles of activity
(Fig.~\ref{fig:num_tweets}). Consequently, we group the data in weekly
bins.  Given that the date of the original message and the date of its
retweet are not always the same, retweets are labelled by the date of
the original posting.

\begin{figure}[t]
  \centerline{\includegraphics[width=0.35\textwidth]{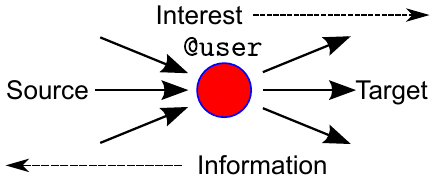}}
  \caption{ In retweet networks, nodes that are exposed to content can
    ``pass it on'' to their followers. The flow of data goes in the
    opposite direction to the interest of the users, where interest is
    represented by arrows which indicate the direction of attention.}
  \label{fig:rt_netwk}
\end{figure}

\begin{figure}[t]
  \centerline{\includegraphics[width=0.45\textwidth]{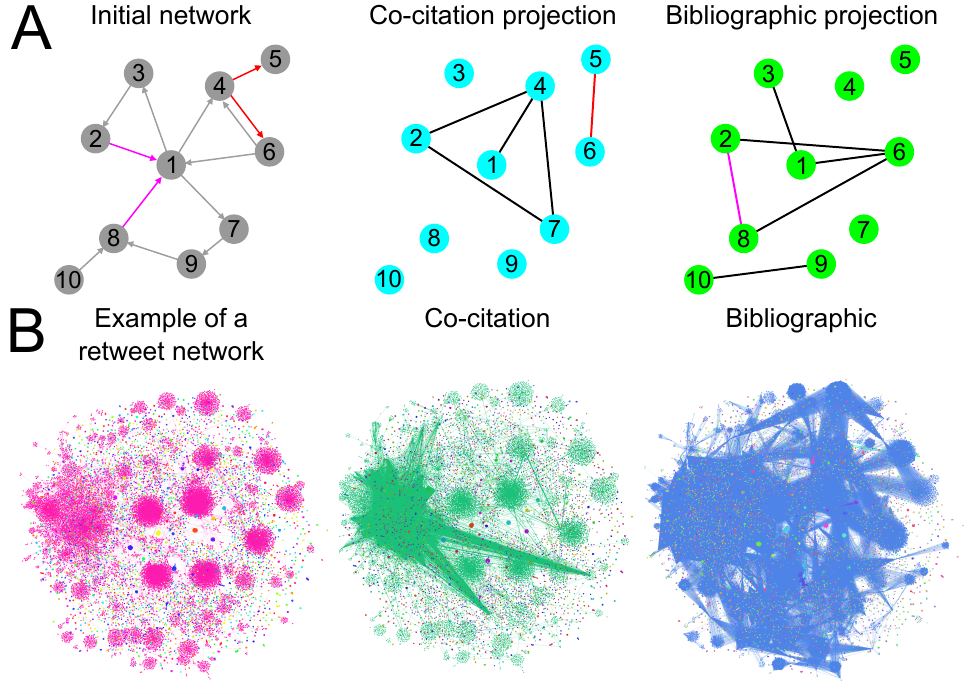}}
  \caption{ {\bf A}: Projections of a retweet network. The original
    retweet network is directed. The co-citation projection joins
    nodes who are the targets of the interest of other nodes. Nodes 5
    and 6 have both been rewteeted by node 4, hence they are connected
    in the co-citation projection (edges in red). In turn, nodes 2 and
    8 have both retweeted statuses by node 1, which is why they are
    connected in the bibliographic projection (edges in magenta). Both
    of the projections have several disconnected components. {\bf B}:
    Example of retweet, co-citation and bibliographic networks
    constructed from one week's worth of interactions in our
    data. There are approximately 44,000 nodes, each coloured
    according to their weakly-connected component in each network.}
  \label{fig:projections}
\end{figure}

\paragraph*{\textbf{Retweet network:}}
For each week $w$, we construct a directed, weighted \textit{retweet
  network} with $N(w)$ nodes, corresponding to all the users who
participated in a retweet event either actively (by retweeting) or
passively (by having their statuses retweeted by someone else).  A
directed connection (edge or arc) exists between two nodes if the
source node has retweeted the status of the target node, and the
weight of the edge corresponds to the number of retweets (see
Fig.~\ref{fig:rt_netwk}), i.e., we define the $N(w) \times N(w)$
adjacency matrix of the retweet network as $\mathbf{A}(w)$, where the
element $A_{ij}(w)$ records the number of retweets from $i$ to $j$
over week $w$.  For a given node, the in-degree corresponds to the
number of users who have retweeted statuses of that node, and the
out-degree corresponds to the number of users whose status that node
has retweeted.  In the example network from
Fig.~\ref{fig:projections}A, node 1 has retweeted statuses of nodes 3,
4 and 7, and has had nodes 2, 6 and 8 retweet some of her own
(i.e. the in-and out-degree of node 1 are both three).  A
weakly-connected component is a group of nodes that are all mutually
reachable if we ignore the direction of the edges. For example, the
initial network in Fig.~\ref{fig:projections}A consists of only one
weakly-connected component, but its co-citation and bibliographic
projections (defined below) have six and five components respectively.

In a retweet network, the direction of the edge signifies an explicit
declaration of interest (i.e. source nodes find certain messages
worth passing along to their followers) and information flows in the
opposite direction (i.e. from the target nodes), as shown in
Fig.~\ref{fig:rt_netwk}.  Such retweet networks have a distinct
structure, with many weakly-connected components and an abundance of
`star' motifs, in which a highly-retweeted node (with high in-degree)
is surrounded by many nodes that point almost exclusively to the star
node.  Figure~\ref{fig:projections}B shows a typical network
for one week in our data with 34,000
nodes and about 4,000 weakly-connected components. Over 50\% of
the nodes belong to a giant connected component (nodes in purple)
whereas each of the smaller components contain less that 1\% of the
nodes.

A fundamental feature of retweet networks is the fact that they are
directed (they may contain extreme asymmetry of interest, `leaders'
and `followers' and other roles).  Directionality entails
additional computational challenges, which is why it is often
neglected. However, ignoring directionality in such settings destroys
valuable information and can severely affect the results and their
interpretation~\cite{Beguerisse2014}.  Our analysis below takes the
directed and temporal nature of the network into full account.

\paragraph*{\textbf{Co-citation and bibliographic projections
    of the retweet network:}} To understand not only who drives
conversations and produces influential content, but also which users
are instrumental in the dissemination of content, we study two useful
networks derived from the original retweet network: the {\it
  co-citation} and {\it bibliographic} projections~\cite{Newman2010}.
These projections reflect the asymmetry of the original adjacency
retweet matrix $\mathbf{A}(w)$, and provide (a) information about the
shared interests of users who have retweeted tweets by the same
authors (bibliographic) and (b) information about whose tweets
generate responses from the same users (co-citation).

\begin{figure}[t]
  \centerline{\includegraphics[width=0.45\textwidth]{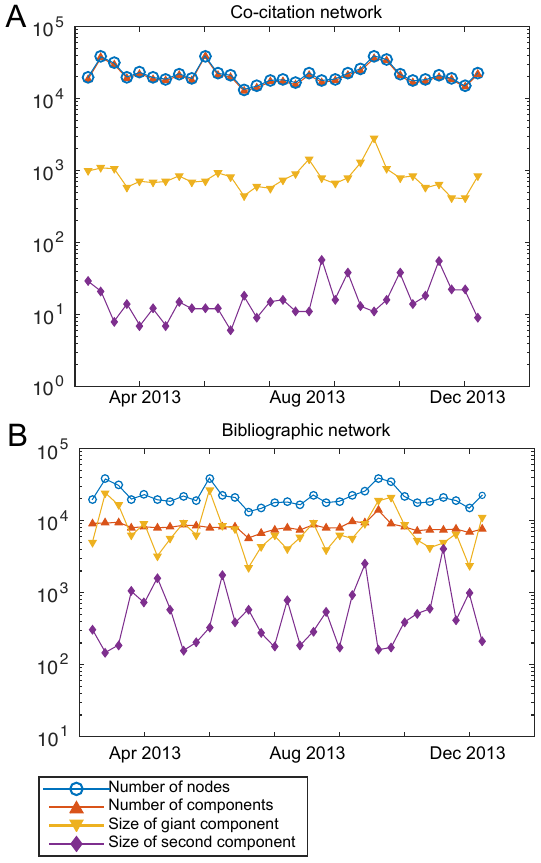}}
  \caption{{\bf A}: Number of nodes, number of components and size of
    the two largest connected components in the co-citation projection
    through the observation period.  {\bf B}: Same measurements for
    the bibliographic projection.}
  \label{fig:components}
\end{figure}

The \textit{co-citation network projection} of $\mathbf{A}(w)$ is an
undirected, weighted network defined on the same set of $N(w)$ nodes as the
original retweet network with adjacency matrix $\mathbf{C}(w)$:
\begin{equation}
  \mathbf{C}(w) = \mathbf{A}(w)^T\mathbf{A}(w).
\end{equation}
An edge of the co-citation network $\mathbf{C}_{i,j}(w)$ exists when
at least one user has retweeted messages from both users $i$ and $j$
during week $w$.  Hence this network links authors of tweets that
elicit interest in users, even though they themselves may
belong to different spheres of interest and activity
(Fig.~\ref{fig:projections}).  The number of disconnected nodes in the
co-citation network is typically large (Fig.~\ref{fig:components}A).
For example, all nodes whose in-neighbours have a maximum out-degree
equal to 1 in the retweet network will be isolated in the co-citation
projection (Fig.~\ref{fig:projections}). 

The \textit{bibliographic network projection} is the converse of the
the co-citation network. It is also built from the original retweet
network $\mathbf{A}(w)$ and has adjacency matrix
\begin{equation}
  \mathbf{B}(w) = \mathbf{A}(w)\mathbf{A}(w)^T.
\end{equation}
This projection links pairs of users who have retweeted messages
posted by at least one author in common; in other words, the
bibliographic projection connects users who share interests.  In a
retweet networks context the size of the largest component in the
bibliographic projection is typically larger than in the co-citation
projection (Figs.~\ref{fig:projections}B and~\ref{fig:components});
this is due to the fact that, on average, Twitter users retweet more
than they are rewteeted.

Note that all nodes in the original retweet network exist in both
projections.  However, depending on their connectivity (induced by the
interest elicited or the interest bestowed) they may appear isolated
in one or both of the projections.  Figure~\ref{fig:projections} shows
an illustration of the construction of the co-citation and
bibliographic projections in a simple example, as well as the retweet,
co-citation and bibliographic networks compiled from a typical week in
the data.

\paragraph*{\textbf{Follower network:}}

In addition to the retweet network, we also constructed a {\it
  follower network} for a small subset of users of interest (see
Sec.~\ref{sec:auth_follower} below). In this network a directed
connection between two nodes exists when the source node {\it follows}
the target node on Twitter (i.e., is subscribed to the target's
content). The interpretation of the follower network is similar to the
retweet network (Fig.~\ref{fig:rt_netwk}) in the sense that a directed
connection is a declaration of interest, with content flowing in the
opposite direction~\cite{Beguerisse2014}. One key difference between
retweet and follower networks is that connections in the retweet
network can be understood as localised expressions of interest
associated to a specific tweet, whereas in follower networks a
connection may be understood as a more general (and stable) expression
of interest.

\subsection{Computation of the hub and authority scores over time}
\label{sec:hubauth}

Hub and authority scores provide a useful tool to analyse directed
networks~\cite{Kleinberg1999}.  Intuitively, a {\it hub} is a node
whose outgoing connections point to important nodes in the retweet
network (authorities); conversely, an {\it authority} is a node with
many incoming connections from hubs.  The entries of the lead
eigenvectors of the adjacency matrices of the co-citation and
bibliographic projections $\mathbf{C}(w)$ and $\mathbf{B}(w)$
correspond to the \textit{authority and hub centrality scores} of the
nodes during week $w$.  A node may rank highly as hub, authority, both
or none, as we will show below.

Henceforth, for notational simplicity, we drop the $w$ from
the matrices $ \mathbf{A}(w), \mathbf{C}(w)$ and $\mathbf{B}(w)$,
but the dependence on $w$ remains implicit.

An alternative way to understand hub and authority scores is through
the singular value decomposition (SVD) of $\mathbf{A}$: $\mathbf{A} =
\mathbf{U}\boldsymbol{\Sigma}\mathbf{V}^T,$ where $\mathbf{U}$ and
$\mathbf{V}^T$ are unitary matrices containing the left and right
singular vectors, respectively, and
$\boldsymbol{\Sigma}=\mathrm{diag}(\sigma_i)$ is the diagonal matrix
of the $N$ singular values.  The entries of the leading left and right
singular vectors of $\mathbf{A}$ correspond to the hub and authority
scores, respectively~\cite{Ding2004}.

As explained above, the bibliographic and co-citation networks contain
several disconnected components
(Figs.~\ref{fig:projections}~and~\ref{fig:components}).  The largest
connected component in a typical co-citation network in our dataset
contains about 10-15\% of the nodes, yet it is responsible for
producing about 50\% of messages that have at least one retweet.  This
observation is consistent with Ref.~\cite{Wu2011} which reports that
about 50\% of URLs on Twitter are posted by a small minority of `elite
users'.  On the other hand, since
the number of users who retweet is much larger than those whose
messages are retweeted, the bibliographic projection has a much
larger giant connected component, with about 75\% of the nodes.

Given the fragmented nature of the projection networks, we compute
hub/authority scores weight-averaged over all components as
follows. Let the co-citation network for a given week $\mathbf{C}$
have $K$ weakly-connected components (excluding components of size 1),
which we label with the index $k$ in order of decreasing size (i.e.,
$k=1$ corresponds to the largest and $k=K$ to the smallest
component). Therefore, $\mathbf{C}=\bigoplus_{k=1}^K\mathbf{C}_k$,
where $\mathbf{C}_k$ is the
$N_k \times N_k$ adjacency matrix of the $k$-th component.
For each component
$k$, we compute the authority score, $\mathbf{v}_k$:
$$\mathbf{C}_k\mathbf{v}_k=\lambda^{\max}_{k}\, \mathbf{v}_k,$$ where
$\lambda^{\max}_{k}$ is the largest eigenvalue and $\mathbf{v}_k$ is
normalised such that all its entries add up to 1 (i.e.,
$\norm{\mathbf{v}_k}_1=1$). The authority score of the $i$-th user in
component $k$ is $\mathbf{v}_k(i)$.  We then aggregate the authority
scores of all components into the weight-averaged vector $\mathbf{v} =
\left[
  (N_1/N)\,\mathbf{v}_1^T,\,\dots,\,(N_K/N)\,\mathbf{v}_K^T\right]^T$.
The score for each node is weighed according to the size of the
component in which it is found, thus ensuring that the scores are
comparable and that we do not discard information from the smaller
components.

To extract the hub scores in each week we follow the same procedure on
matrix $\mathbf{B}$.

\subsection{Topic extraction}

Topic extraction from the data proceeds in several steps. First, the
tweets in the dataset are grouped in weekly bins (as for the networks
constructed above).  For each week, the raw text is pre-processed and
a word co-occurrence network is created. The probability that the
tweets in each bin belong to different topics is computed using
techniques from textual analysis and community detection for
graphs. Note that topics are extracted using only the original
tweets---retweets are excluded because they do not add any new topical
information.  The details of each of the steps for topic
extraction are as follows:

\renewcommand{\arraystretch}{1.1}
\begin{table*}[t]
  \smaller
  \begin{tabular}{|p{0.5\textwidth}|p{0.5\textwidth}|}
    \hline
    {\normalsize \bf Raw tweet} & {\normalsize \bf Processed tweet} \\
    \hline
    US FDA approves Johnson \& Johnson diabetes drug, canagliflozin
    - Reuters http://t.co/pKYCbqiVAZ \#health
    & us fda approv johnson johnson drug canagliflozin reuter health \\
    \hline
    SAVOR-TIMI 53 sets new standard for cardiovascular outcome
    trials in diabetes http://t.co/j9jkB1RagE \#pharma
    &  savortimi set standard cardiovascular outcom trial pharma \\
    \hline
    RESEARCH AND MARKETS: Type 2 Diabetes - Pipeline Review, H2 2013
    http://t.co/5afh5ypDTl
    & research market type2 pipelin review h2 \\
    \hline
    Last Dec my son (then 4) was diagnosed with type 1
    Diabetes. JDRF UK do excellent research into prevention, lifestyle
    and cure. \#walkforcure
    &  last dec diagnos type1 jdrf uk excel research prevent
                   lifestyl cure walkforcur \\
    \hline
  \end{tabular}
  \caption{Examples of `raw tweets' and their processed version prior
    to topic extraction.}
  \label{tab:testproc}
\end{table*}

\paragraph{\textbf{Text pre-processing:}}
Prior to topic extraction, the text of each
tweet is processed in the following way:
\begin{itemize}
\item Convert words to lowercase, e.g., the terms
  `diabetes', `Diabetes' and `DIABETES' are all processed as
  the same word.
\item Remove all punctuation signs and non-alphanumeric characters, i.e.,
  we compare tweets based only on the words they contain.
 
\item Replace collocations for specific terms (e.g., `Type 2' or `T2D'
  are replaced by `type2') in order to homogenise terms that are
  known to refer to the same concept.
\item Remove stop-words and special words (articles, conjunctions,
  etc)~\cite{Manning2008}, numbers and vestigial URLs, which bear no
  topical information.
\item Stem the text~\cite{Porter2009}. This step strips suffixes so that
  related words are mapped to the same stem, e.g.,
  `house', `houses', `housing' are replaced by `hous'.
\end{itemize}
The result of this pre-processing step is illustrated in
Table~\ref{tab:testproc}, which contains examples of `raw tweets' and
their processed version.

\paragraph{\textbf{Topic extraction from word co-occurrence graphs:}}

\begin{figure}[t]
  \centerline{\includegraphics[width=0.45\textwidth]{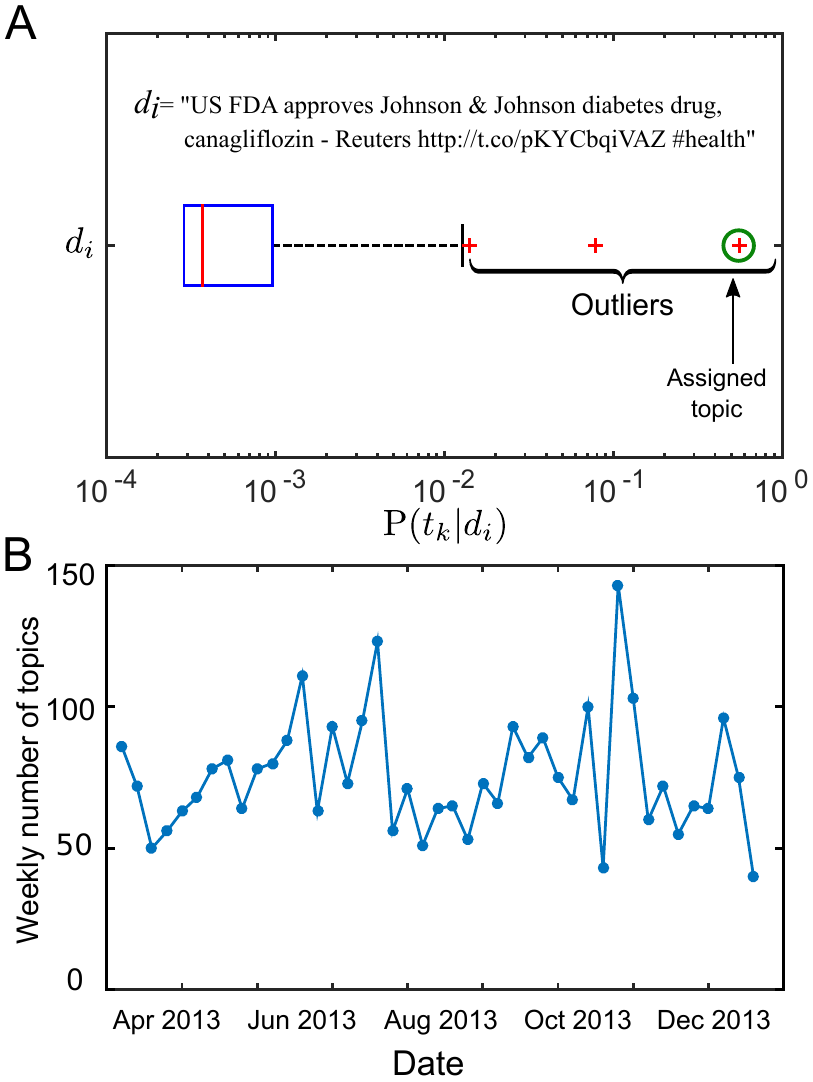}}
  \caption{ {\bf A}: Box-plot of the posterior probability of a
    specific tweet $d_i$ after LDA optimisation.  The red line
    indicates the median LDA assignment score for this tweet, the box
    indicates the 25th and 75th percentiles, and the whiskers mark the
    interpercentile range (IPR, see text). There are three topics (red
    crosses) that lie outside the interpercentile range, the one
    furthest to the right is designated as the topic of the tweet.
    This tweet belongs to a topic which covers the FDA's approval of a
    diabetes drug in late March 2013. The topic belongs to a theme the
    `Commercial' thematic group in Table~\ref{tab:topical_tweets}.
    {\bf B}: The number of topics extracted varies for each week with
    an average of 75.}
  \label{fig:boxplot}
\end{figure}

After text pre-processing, we extract the topics from the tweets in
each weekly bin in the following steps:

\begin{enumerate}
\item Following Lancichinetti \textit{et al}~\cite{Lancichinetti2015},
  we create a word adjacency graph for each weekly bin. The nodes in
  the graph are words, and edges indicate two words that co-appear in
  tweets with higher probability than one would expect at random
  (i.e., when the probability of such an edge in a random network with
  the same degree sequence is less than $0.05$).

\item We analyse the word co-occurrence graph using Markov
  Stability~\cite{Delvenne2010,Delvenne2013} (in contrast with the use
  of Infomap in Ref.~\cite{Lancichinetti2015}) to extract relevant
  communities (or groups) of words that co-appear in tweets more
  consistently than with words outside of their own group, and to consider
  groupings of different granularity~\cite{Beguerisse2014,Lambiotte2014}.

\item The communities of words are used as the input for the {\it
    Latent Dirichlet Allocation} (LDA) topic extraction
  method~\cite{Blei2003}. Hence, the communities obtained with Markov Stability
  provide an \emph{initial} guess for the topics to which a tweet is
  likely to belong. Specifically, we assume that each document (tweet)
  $d_i$ belongs to a topic (community) $t_k$ with a prior probability
  $P_0(d_i | t_k)$ that is proportional to the number of words that
  document $d_i$ has in topic $t_k$.  LDA then proceeds iteratively to produce 
  the posterior probability that document $d_i$ belongs to each
  topic $t_k$: $P(t_k | d_i)$. LDA also produces $P(h_j|t_k)$, the probability
  of finding word $h_j$ given topic $t_k$.

\item We make the assumption that, given their brevity, each tweet
  can only belong to \textit{at most one topic}.  This entails
  collecting only the assignments of topics to documents with
  abnormally large values (i.e., the outliers) of $P(t_k | d_i)$, and
  discarding all other values.  A tweet is assigned to a topic if
  $P(t_k|d_i) > \mathrm{pc}(99) + \mathrm{IPR}$, where $\mathrm{pc}(99)$
  is the 99th percentile of the distribution and IPR is the {\it
    inter-percentile range}: $\mathrm{IPR} =
  \mathrm{pc}(99)-\mathrm{pc}(1)$.  Figure~\ref{fig:boxplot}A shows an
  example of this criterion where a tweet about the FDA's
  approval of a diabetes drug is assigned to a topic that contains
  almost 2,000 tweets about this event.
\end{enumerate}

This method for topic extraction is applied separately to each of the
43 weeks in our dataset.  Figure~\ref{fig:boxplot}B shows the number of
topics in each week.  The average number of topics per week is 75.

\subsection{Themes from topics:}

In order to identify the dominant themes in the dataset, we applied
\textit{open thematic coding}~\cite{Braun2006} to the topics obtained
in the previous section, for a random selection of
weeks. In total, we analysed in-depth 290 topics (from weeks 34, 39,
44 and 50 of 2013, and week 2 of 2014) containing 63,000 processed
tweets with identified topic.

The analysis consisted of reading all the tweets in each topic, and
assigning a label to the topic according to the dominant theme
present.  In general, one or two themes clearly dominated each topic.
In some cases, tweets in one topic were too mixed to identify one
single theme; in this case, we did not assign a theme label but we read
the tweets to ensure that no new themes emerged within the topic.  As
per standard practice, we continued this process until saturation;
that is, until no new themes emerged from the evaluation of topics,
and the repetition of existing theme labels was
constant~\cite{Bowen2008}.  This process was repeated twice by the
same researcher to ensure consistency.  For ease of representation, we then
grouped the themes identified through open thematic coding into thematic
groups, as seen in Table~\ref{tab:topical_tweets}.

In summary, our analysis of content progressed as follows: first, we
automatically extracted weekly {\it topics} from the tweets; second,
we manually applied open thematic coding until saturation to obtain a
list of coded \textit{themes}; and third, we manually placed these
themes into \textit{thematic groups}.

In order to learn which themes dominated in tweets by the most
influential accounts, we also analysed the tweets by the top
authority nodes. Here, tweets belonging to each user were
coded according to themes identified in the thematic analysis.

\subsection{Interest communities in the follower network of authorities:}

To complement the textual analysis above, we also analysed the 
community structure of the \textit{follower network} of the
1,000 accounts with the highest cumulative authority score over the
observation period. The giant connected component of this network comprises 880
nodes, with the rest being either isolated or suspended. To extract
the communities in the networks, we used the Markov Stability
community detection framework as described in
Refs.~\cite{Beguerisse2014} and~\cite{Amor2015}; we found a robust
partition of the network into seven communities of different size. We
labelled the communities according to the type of
accounts that each contains (e.g., health agencies, diabetes advocacy,
research scientists, and so on).

\section{Results}

\subsection{The `what' of diabetes on Twitter}

As described above, the tweets on our dataset were first automatically
grouped into topics, and their content was then manually assigned
themes using open thematic coding. These themes fall into one of four
broad thematic groups: health information, news, social interaction,
and commercial.  For tweets in these groups, although the content and
messages changed from week to week, the themes remained constant. In
addition to these four thematic groups, we identified a fifth group of
tweets with consistent and recurrent content.  In this distinct group,
the specific content was repeated consistently across the data sample
(i.e., the tweets did not change from week to week).  In
Table~\ref{tab:topical_tweets}, we present the thematic groups and the
themes within each group. Figure~\ref{fig:wordclouds} contains a
visual summary of the thematic groups as 'word clouds' with the 200
top words in each thematic group (i.e., the words most likely to
appear in the topics), and `word clouds' with the usernames (scaled to
size of number of tweets) of the 200 most active users (available also
as a Supplementary spreadsheet (see Data Statement).  Note that the
term {\it diabetes} has been removed from the analysis because it
appears in all of the tweets. It is important to remark that the most
active users (e.g., those who tweet the most) are not necessarily the
most important, or `central' in the different networks, from the point
of view of information generation.

We now discuss in more detail the themes, topics and users found in
each of the five thematic groups.

\renewcommand{\arraystretch}{1.1}
\begin{table*}[tp]
  \smaller
  \begin{tabular}{|p{0.2\textwidth}|p{0.8\textwidth}|}
    \hline
    {\normalsize \bf Thematic group} & {\normalsize \bf Theme} \\
    \hline
    \multirow{8}{*}{\multititle{Health information} }
    & Public health messages  \\
    & Links to articles, blogs and studies about risks,
      treatment and cure \\
    & Population health fears \\
    & Publicity about outreach and awareness events and activities \\
    & Advice about diabetes management and diagnosis \\
    & Lifestyle, diet and cookery tips, news and links \\
    & Life stories and experiences (some for marketing purposes)  \\
    & Dangers of sugar, sugar replacements and/or soda  \\
    \hline
    \multirow{4}{*}{\multititle{News}}
    & Headline links to particular `breakthrough' studies or technologies\\
    & Celebrity news \\
    & General news articles about diabetic people or pets \\
    & News relating to the pharmaceutical industry and the economy \\
    \hline
    \multirow{6}{*}{\multititle{Social Interaction}}
    & Users joking about how what they have eaten is likely to give
      them diabetes \\
    & Chatter and social interchanges that include mentions of diabetes \\
    & Everyday experiences of diabetes\\
    & Stigmatising comments\\
    & Banter and sexual innuendo and humour relating to sweetness
      and diabetes \\
    \hline
    \multirow{4}{*}{\multititle{Commercial}}
    & Advertisements for jobs in the pharmaceutical and care industries \\
    & Marketing for a specific product, app, treatment, event or service\\
    & Pharmaceutical, health industry and stockmarket updates and FDA
      approvals\\
    & Sale of diabetes drugs, diets or treatment products online\\
    \hline \hline
    \multirow{6}{*}{\multititle{Recurrent content}}
    & Song lyric: `All the time' by Jeremih~\cite{Jeremih2013}\\
    & Song lyric: `I'm still happy' by Boosie Badazz~\cite{Boosie2010}\\
    & Viral `fact': Alcohol reduces diabetes risk\\
    & Viral `fact': Urine tasting \\
    & The mathematics joke \\
    \hline
  \end{tabular}
  \caption{Thematic groups and associated themes obtained through open
    coding applied to the topics detected in the tweets of the
    collected dataset.}
  \label{tab:topical_tweets}
\end{table*}

\subsubsection{Health information tweets}

One of the largest thematic groups consists of health information,
research findings, recommendations, advice and warnings, which are all
abundantly tweeted and retweeted.  Figure~\ref{fig:wordclouds}A
contains a word-cloud with the 200 most probable words in the topics
of this group (the total number of distinct terms is 29,647 in 90
topics from open thematic coding).  The top 10 words in this thematic
group are: {\it risk, type2, disea (disease, diseased), heart,
  research, month, obe (obese, obesity), fruit, news, awar (aware,
  awareness)}. Such terms are
typical of tweets that fall within this broad thematic group, as seen
in Figure~\ref{fig:wordclouds}A and Supplementary Spreadsheet, as well
as in Table~\ref{tab:healthtweets} in the Appendix containing specific
examples of tweets. The tweets in this group include information about
diabetes (its causes, treatments and cures); technologies and
pharmaceutical products that can be used for managing it; as well as
risks associated with the disease.  Other health-related messages
include publicity about outreach and awareness events, activities and
information.  These individual messages are typically not long-lived,
and are only visible at the top of a user's timeline for short periods
of time~\cite{Gleeson2015}.  In general, there is a high turnover in
the content that each user is exposed to, even though many messages
(e.g., those from newspapers and online media) are posted multiple
times. One of the themes that is less variable in this
regard concerns the dangers of sugar, sugar replacements and/or sodas. 
Its frequency is unsurprising given the nature of diabetes as
a problem of blood sugar regulation.  It is worth mentioning that the
largest spike of activity in our data (in November 2013, see
Fig.~\ref{fig:num_tweets}) is in part due to a surge in tweets about
World Diabetes Day on 14 November 2013.

The health information in the collected sample does not appear to be
directly or specifically associated with health promotion
groups. Instead, such tweets are posted by users with different claims
to expertise: individuals who have experienced diabetes; personal
trainers marketing their services; companies selling lifestyle
products or services; and other users with an apparent interest in
diabetes, cookery  and healthy eating.
These tweets include advice about diabetes management and diagnosis,
cookery and diet tips, life stories and experiences, and links to
articles about new treatments or promising cures. Some tweets make
claims about `curing' diabetes, or offer natural or
`miracle' treatments. Advice of this nature appears to be
authoritative in tone and language, making it difficult to distinguish it
from advice disseminated by official health authorities.  Links
contained in such tweets point to different types of authoritative
sources of content: from people who have first-hand experience of
diabetes, to marketing agencies trying to sell a particular food,
supplement or device, to hospitals attempting to communicate a
specific health message. Home remedies and `miracle' cures appear
alongside health tips and recommendations. Digests and newsletters
specifically containing diabetes-related news (where the authors
gather information from multiple sources around the internet and
supply it to their followers) appear at different moments in time and
mix all such messages together.

\subsubsection{News tweets}

News-related tweets in the collected dataset typically list a headline
of a news article; they sometimes give the first line of the story and
often also provide a weblink to the complete story. These tweets
rarely relate to health promotion or education
messages. Figure~\ref{fig:wordclouds}B contains the 200 most probable
words theme (out of 10,416 in 29 topics).  The top 10 words are: {\it
  type2, risk, fruit, type1, eat, peopl (people, peoples), blueberri
  (blueberry, blueberries), cut, research, juic (juice, juicer,
  juicing, juicy).} Some news-related tweets communicate research
breakthrough studies or technologies, which may be reported with
messages of hope for those who have diabetes, in particular T2. Just
as common in this thematic group are tweets with celebrity news
(especially about celebrities diagnosed with diabetes) and diverse
news about people or pets with diabetes. Another prominent theme in
this group consists of tweets disseminating news headlines about the
pharmaceutical industry and stock market. Diabetes treatment is a
lucrative industry because diabetes is a chronic condition that
requires regular and ongoing treatment (rather than cure), and so the
demand for pharmaceutical products and lifestyle aids is
inelastic~\cite{Simonsen2015}. Furthermore, the number of people with
T2 is projected to increase dramatically in the future as a result of
population ageing and obesity~\cite{Wild2004}, which will further
expand the market. Table~\ref{tab:newstweets} in the Appendix contains
examples of tweets in this category.

\subsubsection{Social interaction and humorous tweets}

\begin{figure}[t]
  \centerline{\includegraphics[width=0.45\textwidth]
    {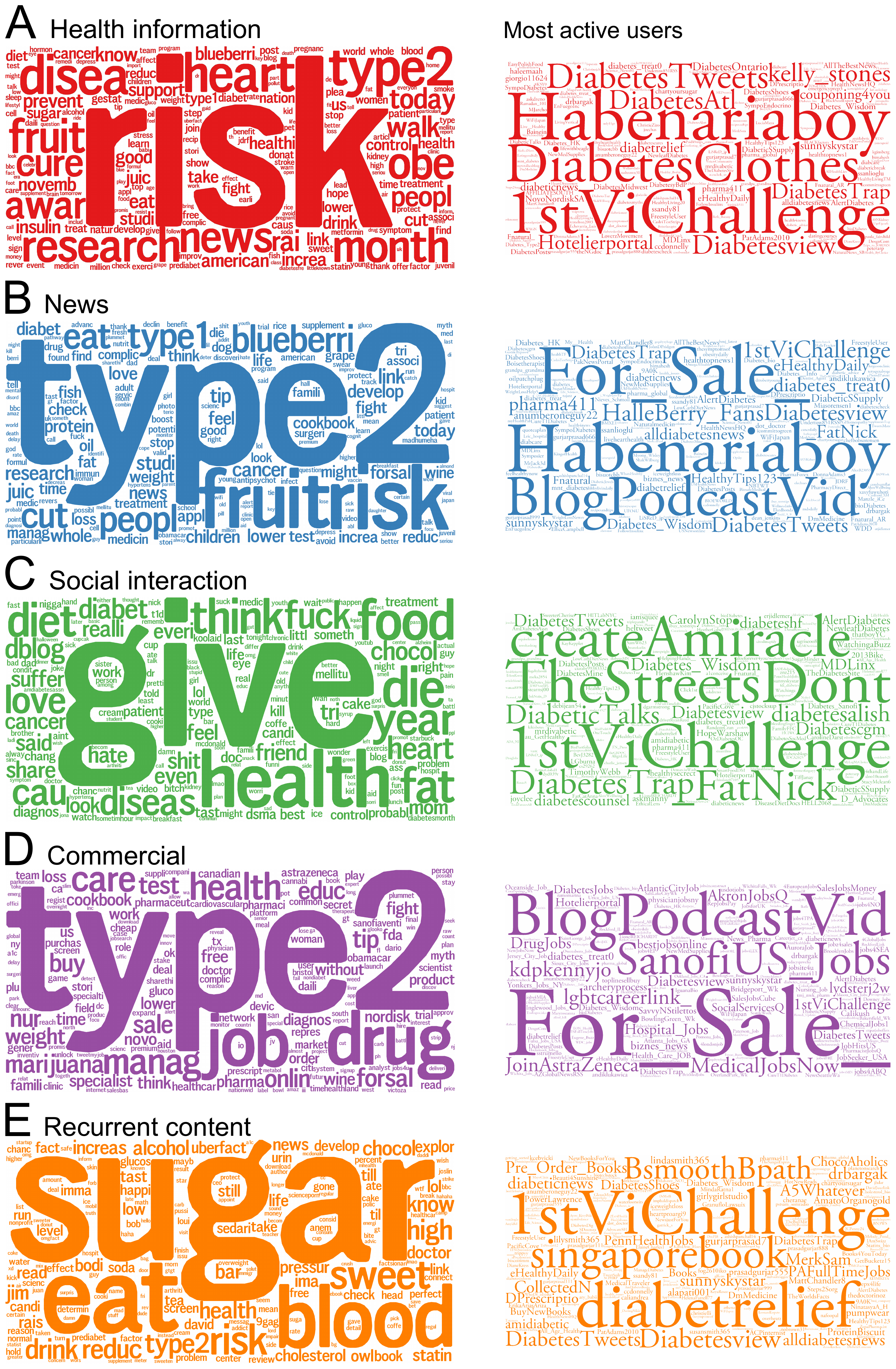}}
  \caption{Top 200 words and most frequent authors in each of the
    thematic groups in Table~\ref{tab:topical_tweets}. The size of
    each word is proportional to its probability of appearing in a
    topic in the group. The size of the username is proportional to
    the number of tweets in the thematic group.}
  \label{fig:wordclouds}
\end{figure}

Twitter is not simply an information-sharing technology, but also a
space that permits (virtual) social interaction~\cite{Kwak2010,
  Gonzalez-Bailon2011}.  Social interaction tweets use language
differently to the thematic groups above: they are typically informal
in tone, their attention to spelling and grammar is limited, and they
often use exclamation marks and punctuation to express fun, laughter,
exasperation, sarcasm, irony, and abuse.  The top 10 words (out of 31,061 in 97
topics) in this group are: {\it give, health, food, die, think, fat,
  year, diet, diseas (disease, diseased), cau (cause, causes,
  causing)}. See Fig.~\ref{fig:wordclouds}C for the theme's
word-cloud.  Users frequently joke about how what they have eaten is
likely to give them diabetes, with a wide variety of sugary foods and
drinks, junk foods and other `unhealthy' options being cited. Such
tweets indicate a level of awareness of dietary guidelines and
diabetes aetiology. Users have conversations and interact about a
diversity of topics in a chatter that is not necessarily directly
related to diabetes but may include references to it.  People who have
diabetes---particularly T1---also talk about their daily experiences
of their bodies, sugar management, and social acceptance or stigma;
such tweets may elicit retweets or messages of support from
others. Some users also talk in terms of ``us'' (with T1) and ``them''
(without). For example, one user talks about T1 as being a `perk' or
feature he/she looks for in a romantic partner:
\begin{displayquote}\smaller
  I haven't stopped thinking about this girl for seriously like...a
  month. AND she has diabetes! \#diabetesperks
\end{displayquote}
Such chatter often elicits retweets and replies, including messages of
support, or appreciation of a joke.

On the other hand, stigmatising comments, especially tweets which
blame diabetic people for bringing the disease on themselves through,
for example, poor diet or lack of physical activity, are abundant in
the dataset. Figure~\ref{fig:wordclouds}C and the Supplementary
Spreadsheet contain numerous examples of profanity and pejoratives.
Faced with such messages, users with T1 diabetes frequently point out
that it is important to differentiate between T1 and T2, insinuating
that while T1 diabetes is not a person's `fault', T2 may well be.
Other tweets include calling other people `diabetic' as an insult and
wishing diabetes upon a person a user does not like.
Table~\ref{tab:SocialInteraction_Humour} in the Appendix contains
examples of social interaction, humorous and stigmatising tweets.

A distinct theme in this category consists of tweets with sexual
innuendo. At their mildest, such tweets refer to boy-band members or
other (often celebrity) `crushes' that are so sweet they are
diabetes-inducing. At their most extreme, tweets joke that others'
bodily fluids and genitals are so sweet they are
diabetes-inducing. These tweets contain weblinks to pornography
websites or other explicit material. Like the jokes discussed earlier,
these tweets might reflect a baseline awareness of the links between
sugar and diabetes.

\subsubsection{Commercial tweets}

As mentioned previously, diabetes is an industry with an attractive
(and expanding) market: it is a chronic condition which is currently
incurable, and it requires constant and regular testing, surveillance
and treatment.  The top 10 words (out of 5,841 words in 20 topics) in
this group are: {\it type2, drug, job, manag (manage, manager), care,
  health, marijuana, sale, test, forsal (for sale)}. See
Fig.~\ref{fig:wordclouds}D for the theme's word-cloud and the most
active users within this topic.  People with diabetes depend on
different technologies, consumables, health services, and
pharmaceutical products.  This commercial dimension of diabetes is
reflected in many Twitter messages. Tweets concerned with stockmarket
listing announcements, FDA approval (e.g., Fig.~\ref{fig:boxplot}A),
court cases, and business recruitment (for example, a company hiring a
new CEO, or advertising for new staff) are common.
Table~\ref{tab:Commercial} in the Appendix contains typical examples
of these types of message.

\subsubsection{Recurrent content}

While the specific content of tweets in different themes varies over
time, there are some tweets whose content appears consistently across
time over our dataset.  The top 10 words (out of 8,731 words in 19
topics) in this theme are: {\it sugar, eat, blood, sweet, risk,
  type2, drink, high, reduc (reduce, reducing, reduction),
  health}. Figure.~\ref{fig:wordclouds}E contains this theme's
word-cloud.  There is a large number of tweets repeating lyrics from
two specific rap songs, both of which contain the term `diabetes' (see
Table~\ref{tab:Recurrent}).  Of these, `All the Time' by
Jeremih~\cite{Jeremih2013} was released in 2013, so it is possible
that the appearance of one line from the song in many tweets in our
sample may be linked to marketing and initial responses to the
song. The lyric repeated from the song relates to the sexual innuendo
discussed previously, so its continued presence may result from it
being used to express a joke rather than as a direct and deliberate
citation (indeed, the song name or artist are rarely mentioned).  The
second song `I'm Still Happy' by Lil Boosie (now Boosie
Badazz)~\cite{Boosie2010} was released on a 2010 digital mixtape,
suggesting that the lyric has been consistently attractive to users
over time.  This lyric is about determination in the face of
challenges; the artist himself has T1 and has talked with his fans
about the challenge of managing his diabetes.


In addition, several specific viral `facts' and jokes are posted
frequently and repeatedly (for example, that tasting urine for
sweetness was a method to detect diabetes in the
past~\cite{Polonsky2012}, or that consuming alcohol moderately has
been reported to reduce diabetes~\cite{Koppes2005}).  These few
specific facts appear intended to entertain or amuse. They retain
almost identical phrasing over time, with sustained popularity.  In
fact, such facts are repeated more consistently over time than other
headlines, health messages and reports.

One of the most prominent instances of recurrent content in
our data corresponds to various versions of a mathematics-themed joke.
A typical instance of this joke is
\begin{displayquote}\smaller
  Math Problems: If Jim has 50 chocolate bars, and eats 45, what does
  he have? Diabetes. Jim has diabetes...
\end{displayquote}
See Table~\ref{tab:Recurrent} in the Appendix for more versions of the
same joke.  This joke appears consistently in our dataset (44,130
times including retweets, Fig.~\ref{fig:num_tweets}). As with other
jokes about certain foods that are linked to diabetes, for this joke
to be amusing or entertaining (which its sheer volume suggests it is
to many) it requires that readers and retweeters have a minimum
awareness of certain foods that might contribute to diabetes.  To
highlight the consistency with which the mathematics joke appears in
our data, we also show on Fig.~\ref{fig:num_tweets} the number of
tweets mentioning the actor Tom Hanks. In October 2013, Tom Hanks
announced he had Type 2 diabetes, which was discussed and shared
widely on Twitter (13,454 tweets in our dataset). In contrast to the
mathematics joke, the number of tweets about Tom Hanks displays an
abrupt `exogenous peak' (evidence of external events affecting the
behaviour of a system~\cite{Sornette2004}) followed by a steady
relaxation to a baseline level.

\subsection{The `who' of diabetes on Twitter}

We now turn our focus to the analysis of the users in the retweet
networks extracted from our dataset. Figure~\ref{fig:wordclouds}
provides a visual illustration of the users that posted more tweets in
the topics contained within each of the thematic groups. Although
helpful, this figure does not provide information about how important
these users are perceived to be in the community, or the impact they
have in the conversation about diabetes. Therefore, in order to
understand who are the key users that influence diabetes-related
content on Twitter over time, we examine the hub and authority scores
of all nodes for each of the weekly retweet networks from June 2013 to
January 2014 (see Section~\ref{sec:hubauth} for a description of the
methodology).  We then examine the content generated by the most
important users (according to their hub and authority scores), and we
finally analyse their follower network (i.e., who {\it follows} whom
within this group of important users).

\subsubsection{Authorities and hubs in the weekly retweet networks}

\paragraph{\textbf{Authority nodes:}}


\begin{figure*}[t]
  \centerline{\includegraphics[width=\textwidth]
    {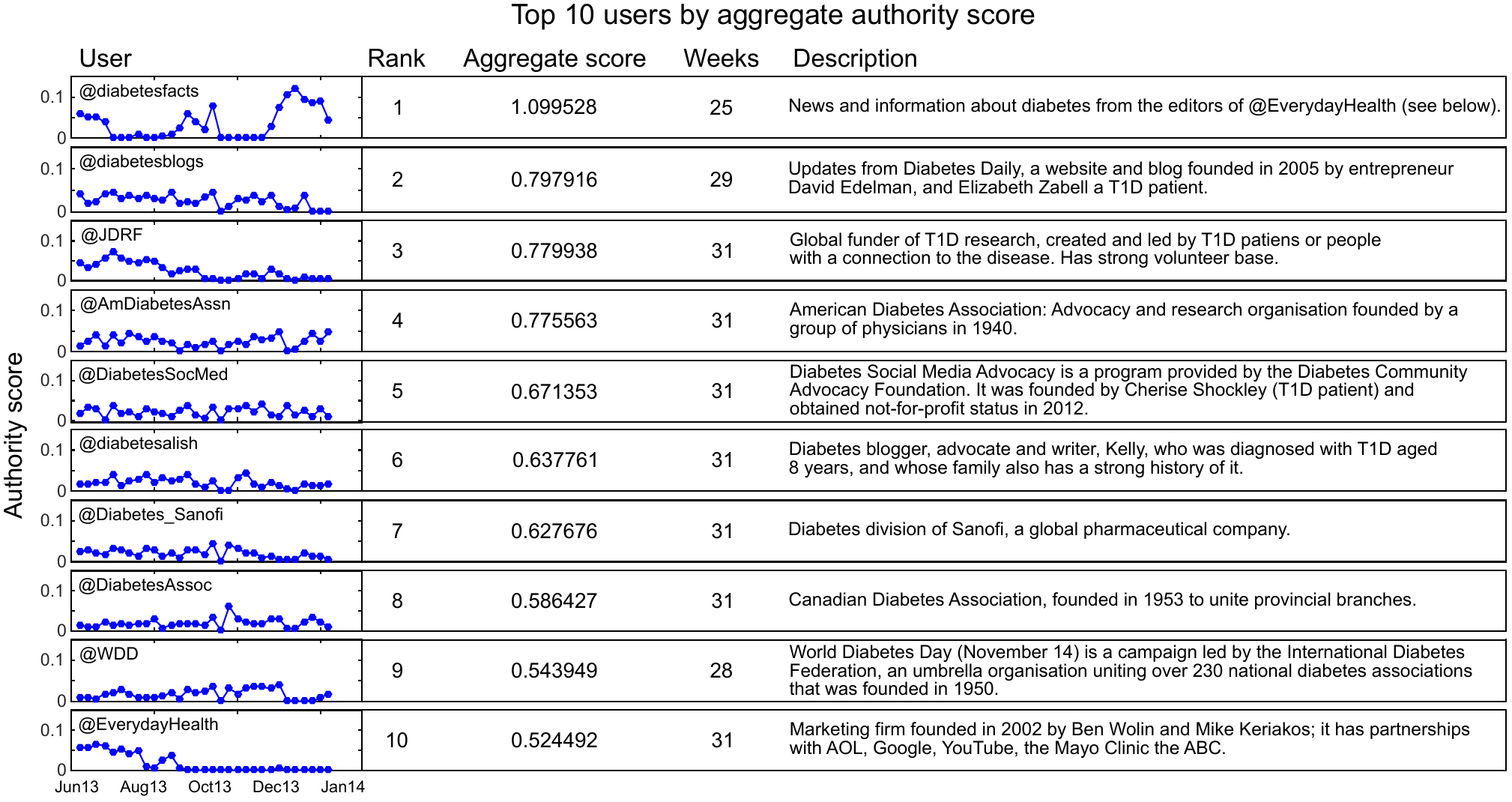}}
  \caption{Top ten users by aggregate authority score, number of weeks
    with nonzero authority score, and brief description.  }
  \label{fig:authorities}
\end{figure*}

Figure~\ref{fig:authorities} shows the score over time for the top ten
authorities ranked by their aggregate weekly score. Of these
authorities, nine are institutional accounts and one belongs to an
individual blogger, Kelly Kunik (@diabetesalish). This blogger is a
diabetes advocate who has had T1 for over 30 years and is involved
with a number of diabetes advocacy organisations. She also
administrates the Diabetesaliciousness
blog~\footnote{\url{http://diabetesaliciousness.blogspot.co.uk/}},
which `spreads diabetes validation through humour, ownership and
advocacy'. Kunik's tweets and blog posts are humorous, casual, and
interpersonal in style.

Three of the institutional accounts belong to stockmarket-listed
commercial ventures (i.e., established with the purpose of generating
profits for shareholders). Of these, two (@diabetesfacts and
@EverydayHealth) belong to Everyday Health Media. This company, which
does not claim to have any specific diabetes expertise, owns and
operates a range of brands. It was founded in 2002 by Ben Wolin (an
entrepreneur who has previously worked for Beliefnet, acquired by News
Corp., PBS, Warner Brothers and Tribune Interactive), and Mike
Keriakos (a media sales expert who began his career at Procter and
Gamble and who built partnerships between Everyday Health and AOL,
Google, YouTube and the ABC).  The other listed company account,
@Diabetes\_Sanofi, belongs to the global pharmaceutical firm Sanofi,
which produces diabetes treatments.

Another three of the top ten authorities correspond to national and
international diabetes associations: @AmDiabetesAssn (the
American Diabetes Association) @DiabetesAssoc (the Canadian
diabetes association) and @WDD (which belongs to the International
Diabetes Federation).  The latter account is established especially to
promote World Diabetes Day (November 14), although it disseminates
diabetes information and messages all year round. These associations
were originally founded by physicians and they focus on advocacy and
research.

The remaining three authorities in the top ten belong to
not-for-profit organisations founded by people who have experienced
T1. These organisations target T1 specifically: one is a funding body
(@JDRF, an organisation previously known as Juvenile Diabetes
Research Foundation), and the other two are blog platforms that host
discussions and disseminate information (@diabetesblogs and
@DiabetesSocMed).

\begin{figure}[t]
  \centerline{\includegraphics[width=0.45\textwidth]{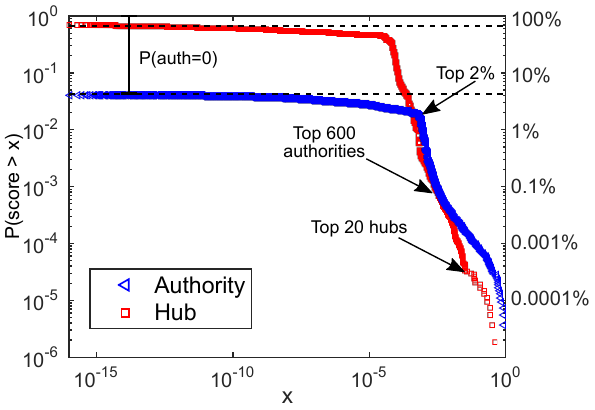}}
  \caption{Cumulative distribution of aggregate hub (red) and
    authority (blue) scores on a
    doubly-logarithmic scale. The horizontal dotted-lines show the
    proportion of nodes with zero aggregate hub ($< 10\%$) and
    authority ($\simeq 95\%$) scores.}
  \label{fig:centralities-cdf}
\end{figure}

Figure~\ref{fig:centralities-cdf} shows the complement of the
cumulative distribution (CDF) (i.e., $1-\mathrm{CDF}$) of the
aggregate authority scores (blue triangles).  The distribution of
authorities shows that over 95\% of users involved in a retweet event
do \emph{not} feature as an authority at all (in our weekly
binning). Users who do attain a non-zero authority score fall into two
categories:
\begin{enumerate}[(i)]
\item users with a sporadic or marginal appearance (i.e., they appear
  in only a handful of weeks and/or not in the giant components of the
  weekly co-citation networks). These users are concentrated in the
  long plateau in Fig.~\ref{fig:centralities-cdf}.
\item users who appear more regularly
usually as part of the weekly giant connected components. These users
constitute the top 2\% in the heavy tail of high scores in
Fig.~\ref{fig:centralities-cdf}.
\end{enumerate}

In the Supplemental Spreadsheet, we
provide the top 1000 accounts by aggregate authority score as a
further resource for research.  As with the top ten users discussed
above, this longer list contains accounts of known commercial, public and
advocacy organisations, as well as medical schools, hospitals,
individual activists and bloggers. In addition, we note the appearance
of academic publishers (@bmj\_latest, ranked 55, @NEJM, ranked 67);
comedians (@ChelcieRice, ranked 28, @SherriEShepherd, ranked 85); news
and media outlets (@medical\_xpress, ranked 40, @foxnewshealth, ranked
82); entertainment (@FoodPorn, ranked 197); and large health-oriented
organisations (@WHO, ranked 88, @NIH, ranked 111), all of which have a
focus much broader than diabetes.  

These accounts disseminate content
and information about diabetes on Twitter efficiently, with messages
that elicit a wide, measurable response through re-tweets and replies
from the general population. The sustained presence of the top
authorities (i.e., the sustained strictly positive authority score
in Fig.~\ref{fig:authorities}) is an indication that these users
consistently produce content that resonates with the general
Twitter-using population.

\paragraph{\textbf{Hub nodes: }}


\begin{figure*}[t]
  \centerline{\includegraphics[width=\textwidth]{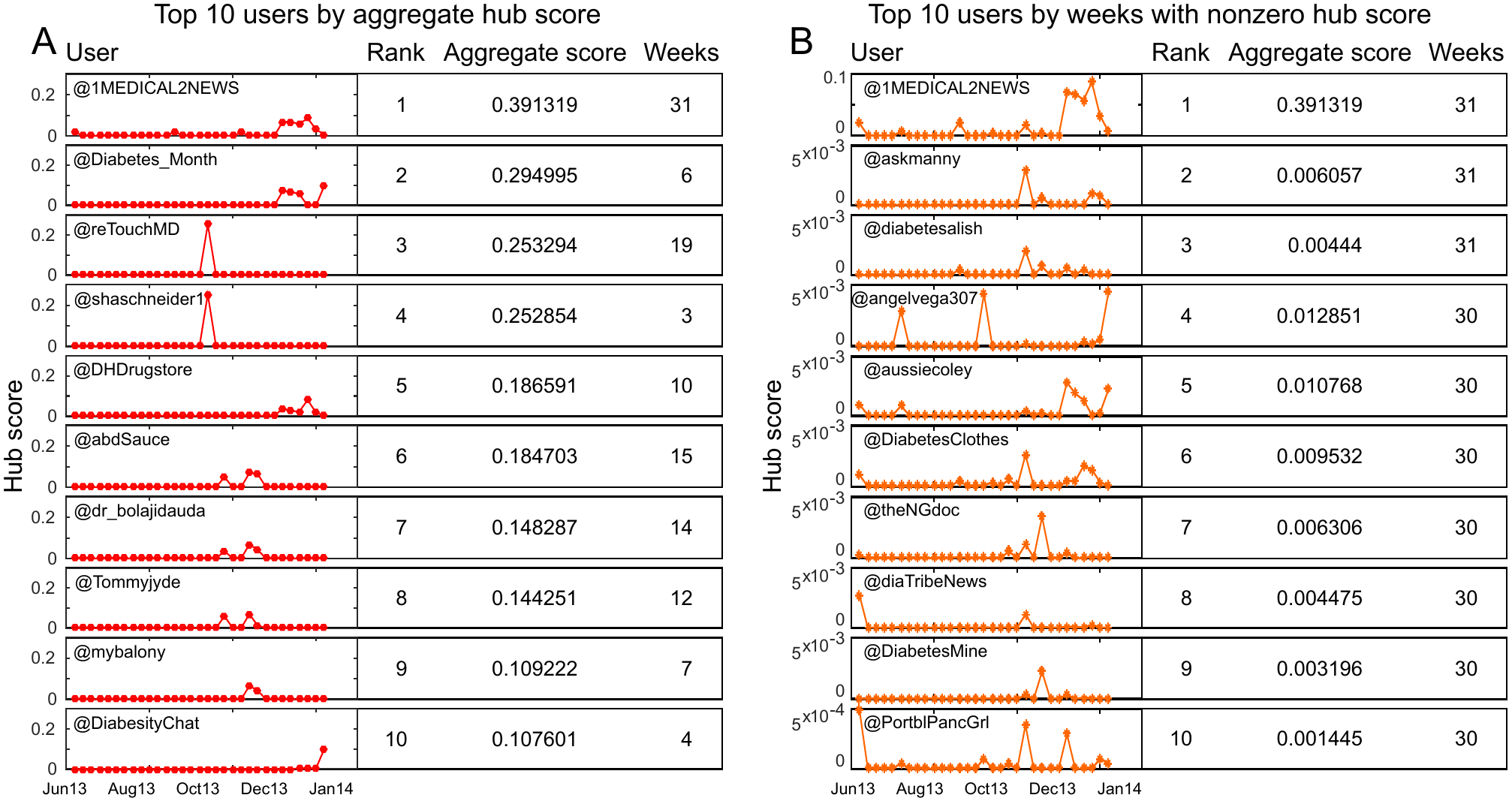}}
  \caption{Two different rankings of hub nodes. {\bf A}: Top ten users
    ranked by their aggregate hub score, along with the number of
    weeks in which they appear with a nonzero hub score. {\bf B}: Top
    ten users ranked by their number of weeks with a nonzero hub
    score.  Note the change in the scale of the $y$-axis. }
  \label{fig:hubs}
\end{figure*}

The top ten accounts by aggregate hub score have a markedly different
character to the authorities, both in behaviour and types of users. In
the context of a retweet network, hubs are connectors; they forward
information generated by authorities across Twitter.  The top accounts
by aggregate hub score over time (see Fig.~\ref{fig:hubs}A and the
Supplemental Spreadsheet) contain: content aggregators (i.e., accounts
that gather internet content for re-use such as @1MEDICAL2NEWS or
@Diabetes\_Month); accounts designed to promote products like
@reTouchMD; automated accounts (e.g., so-called `robots'); and some
individual accounts. Some of these accounts have been suspended
(@shaschneider1); have become inactive (@1MEDICAL2NEWS); or have
seemingly changed hands (@abdSauce) since the data was originally
collected.  

These hub nodes have a highly variable number of followers. For
instance, in August 2016 @Diabetes\_Month had over twenty thousand
followers, whereas @abdSauce had about 50. What all hubs have in
common is a high rate of retweeting and a feeble, unsustained presence
throughout the observation period.  Figure~\ref{fig:hubs}A shows the
scores over time for the top ten hubs based on aggregate score, along
with the number of weeks in which they have a nonzero score.  The data
shows that, in most cases, the high aggregate hub score is the result
of a `one-off' surge.  We thus term these \textit{`intermittent
  hubs'}.

The biographical information provided by these users about who they
are and what they do is sparse, often vague, and difficult to
corroborate.  For example, the account @1MEDICAL2NEWS claims to be a
Dr Richard Billard from Los Angeles, but while the account had a high
level of activity on Twitter in our data (and until August 2014) there
is no other evidence online that this doctor exists. Given the
extremely active Twitter account, it is unusual that this doctor has
absolutely no other online presence. The rate at which this particular
user retweeted (on average, 65 times per day), together with the lack
of any original tweets (or non-retweets), further suggests that it
could have been an automated account assigned an
authoritative-sounding alias.

Figure~\ref{fig:centralities-cdf} shows the complementary CDF
of aggregate hub scores for all users (red squares).  In
contrast with the authority scores, most users ($\sim$ 90\%) have a
non-zero aggregate hub score), and the split between low scoring and
high scoring hubs is more even, with around 50\% of the $\sim$1.2M
users contained in the heavy-tailed part of the distribution.  There
is a distinct bend at the right end of the heavy tailed regime of the
distribution that contains the top 20 users by aggregate hub score, most 
of which have been discussed above.

To obtain a clearer picture of \textit{`persistent hubs'}, we created
an additional activity ranking of hubs by number of weeks in which the
users have a nonzero hub score (Figure~\ref{fig:hubs}B).  The top hub
according to this persistence score is still @1MEDICAL2NEWS, which is
a hub in each of the 31 weeks for which we computed the hub scores.
Another noteworthy appearance is @diabetesalish, who is also a top ten
authority, and who was active as a hub in each week as well. The top
1000 (intermittent and persistent) hub accounts are included in the
Supplemental Spreadsheet as a resource for further research.

\paragraph{\textbf{Authorities vs.\ hubs:}}

\begin{figure}[t]
  \centerline{\includegraphics[width=0.5\textwidth]
    {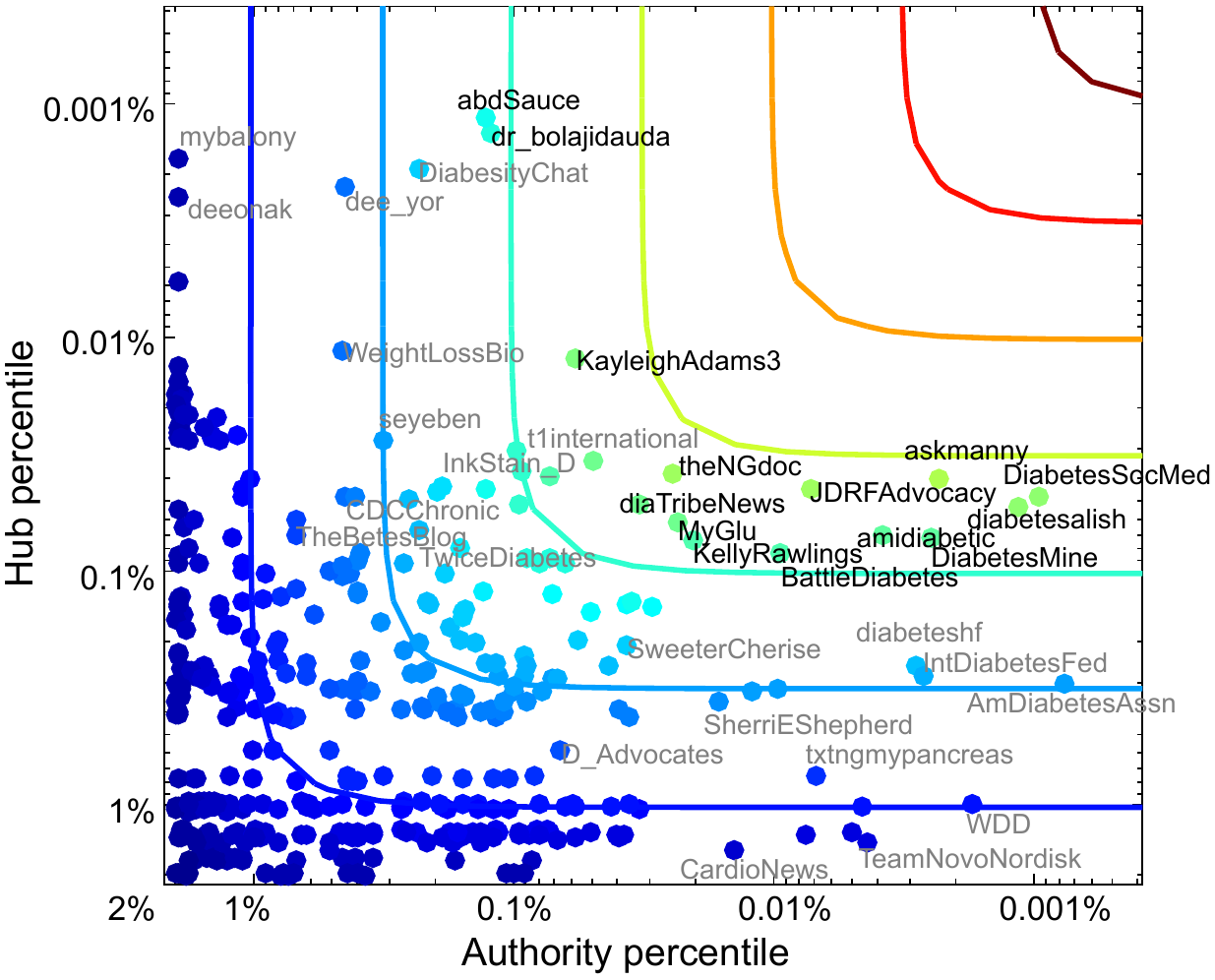}}
  \caption{Twitter accounts with the highest combined aggregate hub
    and authority score (top 2\% on both rankings, on a logarithmic
    scale). The colour of each data point corresponds to the user's
    combined hub/authority score; blue denotes a lower hub or
    authority both (relative to the other accounts on shown here), a
    transition towards red denotes strength on both scores. Shown here
    are the usernames of 35 accounts that score highly both as hubs
    and as authorities; usernames in black are discussed further in
    the text.}
  \label{fig:2D_hub_auth}
\end{figure}

Our results in Figs.~\ref{fig:authorities}, \ref{fig:centralities-cdf}
and~\ref{fig:hubs} show that the behaviour of the top authorities and
hubs in our data is fundamentally different: top authorities tend to
have a more sustained presence throughout the observation period than
top hubs.  We therefore investigate the combined characterisation of
users in terms of joint aggregate hub and authority
scores. Figure~\ref{fig:2D_hub_auth} shows the accounts with high hub
and authority scores (i.e., in the global top 2\% in both rankings).
It is relevant to study users that have a prominent role as
\textit{both hub and authority}.  These include not only the the
blogger @diabetesalish and the advocacy group @DiabetesSocMed, but
also the following users (highlighted in black in
Fig.~\ref{fig:2D_hub_auth}):
\begin{itemize}
\item @JDRFAdvocacy: a US Government-funded organisation which focuses
  on building support for T1D research; the advocacy account of @JDRF;
  (authority rank, hub rank) = (38, 234).
\item @askmanny: Manny Hernandez, a prominent Venezuelan-American
  diabetes advocate who has had diabetes since 2002, and whose
  professional expertise is in social media, technology and health
  (authority rank, hub rank) = (12, 208).
\item @amidiabetic: Diabetes advocate and writer Stuart Wimbles, who
  runs a diabetes information website and reportedly has T1D, although
  there is little additional information about him online; (authority
  rank, hub rank) = (20, 366).
\item @DiabetesMine: A website/blog/newsletter run by Healthline, a
  consumer health information website. The company has a
  large team of staff and a panel of three medical advisors (a
  Professor of Emergency Medicine, a drug information specialist, and
  a pharmacist) who provide insights into user needs;
   (authority rank, hub rank) = (13, 208),
\item @KayleighAdams3: Account currently suspended; 
 (authority rank, hub rank) = (300, 64).
\item @theNGdoc: Nigeria Diabetes Online Community, a non-government
  organisation launched in 2013 to provide empowerment, education and
  support for diabetics in Nigeria. An original idea by Cherise
  Showkley (@DiabetesSocMed), the site is officially recognised by the
  International Diabetes Federation, and two of its Directors are
  medical doctors; (authority rank, hub rank) = (127, 200).
\item @BattleDiabetes: Account of {\tt informationaboutdiabetes.com},
  a website with no details about who it belongs to, but
  which traces to Jill Knapp, a T2 diabetic who is an advocate for the
  American Diabetes Association;
   (authority rank, hub rank) = (49, 442).
\item @KellyRawlings: Diabetes advocate, T1 diabetic, journalist, and
  editor of Diabetes Forecast, a publication of the American Diabetes
  Association; 
   (authority rank, hub rank) =  (107, 387).
\item @diaTribeNews: Account of the diaTribe Foundation, a non-profit
  organisation founded by T1 diabetic Kelly Close to help people
  with diabetes to live better lives. Close was previously a
  financial sector analyst and also runs Close Concerns, a healthcare
  information firm focusing on diabetes and obesity; the Foundation
  Board contains one medical doctor;
   (authority rank, hub rank) = (171, 273).
\item @MyGlu: T1D Exchange, a non-profit
  organisation providing connectivity for people
  whose lives are affected by diabetes. The site's leadership team
  contains a physician and pediatric endocrinologist; but it also
  makes a clear disclaimer that discussions on the site should not be
  a substitute for medical advice;
   (authority rank, hub rank) = (122, 326).
\item @abdSauce: Top hub account, has been taken down or switched
  hands since data were collected; a Pinterest account with the same
  handle belongs to Rasheed Adewole, who works with the Nigeria
  Diabetes Online Community (@theNGdoc);
   (authority rank, hub rank) = (676, 6),
\item @dr\_bolajidauda: Top hub account, no longer active at the time
  of analysis;
   (authority rank, hub rank) = (637, 7).
\end{itemize}
In addition to posting messages that evoke a broad
response (as authorities), these users also engage by retweeting and
replying to messages posted by other users (as hubs).  The empty
upper-right corner in Fig.~\ref{fig:2D_hub_auth} indicates that there
are no accounts at the {\it very top} as both hubs and authorities.

\subsubsection{Extended analysis of authorities: content and relationships.}
\label{sec:auth_follower}

The analysis so far has established the persistence and relevance of a
relatively small number of authorities in the collected retweet
networks. To further our understanding of the group of authorities, we
perform two further analyses.
First, we analyse in detail the thematic content produced by the top ten
authorities. Second, we extract the network of followers within
the top 1,000 authorities and characterise the interest
communities within the network of authorities~\cite{Beguerisse2014,Amor2015}.

\paragraph{\textbf{Topical analysis of authorities:}}

We analyse into which of the topics and themes obtained through 
our earlier analysis of the whole dataset, the tweets from the top 
ten authorities are classified.  All of the top ten authorities post 
messages frequently and consistently
in all the themes listed under the thematic group `Health Information' in
Table~\ref{tab:topical_tweets}.  Some `News' related tweets are also
featured, although these are less common.

Two accounts, @Diabetes\_Sanofi and @diabetesblogs, do not appear to
be dominated by any one of the themes listed, but contain a mixture of
all the `Health Information' themes. Two other accounts, @WDD and
@AmDiabetesAssn participate in themes that are related to outreach and
advocacy activities, events and news.  The not-for-profit organisation
and research funding body @JDRF produces tweets that contain life
stories and experiences of diabetes sufferers more than any other
top-ten authority.

Two accounts, @diabetesfacts and @EverydayHealth (owned by the same
company) focus predominantly on lifestyle and diet-related tips, hints
and advice. Unlike the other authorities, these do not produce
outreach or advocacy messages at all.  Typical messages posted by
these accounts include:
\begin{displayquote}\smaller
  @diabetesfacts: Tips on adjusting your insulin pump
  during exercise from diabetes educator Gary Scheiner @Integ\_Diabetes
  - http://t.co/28Hx8c7PES
\end{displayquote}
\begin{displayquote}\smaller
  @EverydayHealth: Medical costs for people with \#diabetes
  are more than 2x those of people without it.  Are you budgeting for
  diabetes? http://t.co/2ohMOVJvwp
\end{displayquote}
\begin{displayquote}\smaller
  @EverydayHealth: The best beverages to quench
  your thirst with \#diabetes http://t.co/HbR4poFfF5
\end{displayquote}
The vast majority of the tweets from these two accounts provide a link
back to the company's website, which offers articles containing health
and lifestyle advice.

The messages posted by the accounts @diabetesalish and @DiabetesSocMed
are dominated by a mix of social interactions, banter and advocacy.
They participate in news topics, but to a lesser degree than the other
top-ten authorities. Their tone is different to the others: it is
informal and conversational rather than authoritative or
informational. For example:
\begin{displayquote}\smaller
  @DianetesSocMed: Happy Mother's (aunts, fur baby moms, god moms,
  etc.) to all the women in the diabetes community! Have a great day!
\end{displayquote}
and
\begin{displayquote}\smaller
  @diabetesalish: \#dblog:Dear240 \#bgnow: You R but fleeting \&
  temporary, I am permanently fabulous. \#Iwin \#diabetes \#doc \#dsma
  http://t.co/w9iZw83f97
\end{displayquote}
which links to a blog posting about the user's experience with
diabetes. Two other users, \mbox{@diabetesblogs} and
\mbox{@DiabetesAssoc}, also tweet some social and interpersonal
messages.

Two accounts, @diabetesblogs and @diabetesalish occasionally feature
marketing or product promotion messages. Such instances of marketing
are to be expected, as some bloggers generate income by advertising
goods and services, and sponsoring blog advertising in this way is not
regulated by governments. In this case and others, while marketing
might not be made explicit to other users, it is still possible that
ostensibly non-commercial accounts are also practising marketing,
especially accounts owned by stockmarket-listed firms.

The topics where the highest number of top ten authorities converge
are related to advocacy and awareness.  For example, a topic about
Diabetes Blog Week in May 2013 gathered 6 of the top ten authorities:
@diabetesalish, @diabetesblogs, @DiabetesSocMed, @Diabetes\_Sanofi,
@diabetesfacts, and @EverydayHealth.  In other weeks, the top 10
authorities appear together in topics related to promotion of blogs by
diabetics (using the hashtag \#dblogs, which appears in 15,901 tweets
in the data set), and diabetes social media awareness (using the
hashtag \#dsma, which is promoted by @DiabetesSocMed and appears in
10,945 tweets).


\paragraph{\textbf{Analysis of the authority follower network:}}

In order to understand how the users with high authority
scores interact with each other, we have mined Twitter further to
extract and analyse the \textit{follower network} of the top 1000
authorities by aggregate score (see Supplemental Spreadsheet). In this
network, each of the top 1000 authorities is a node and a directed
connection between two nodes indicates that the source node `follows'
the target node on Twitter.

\begin{figure*}[t]
  \centerline{\includegraphics[width=\textwidth]{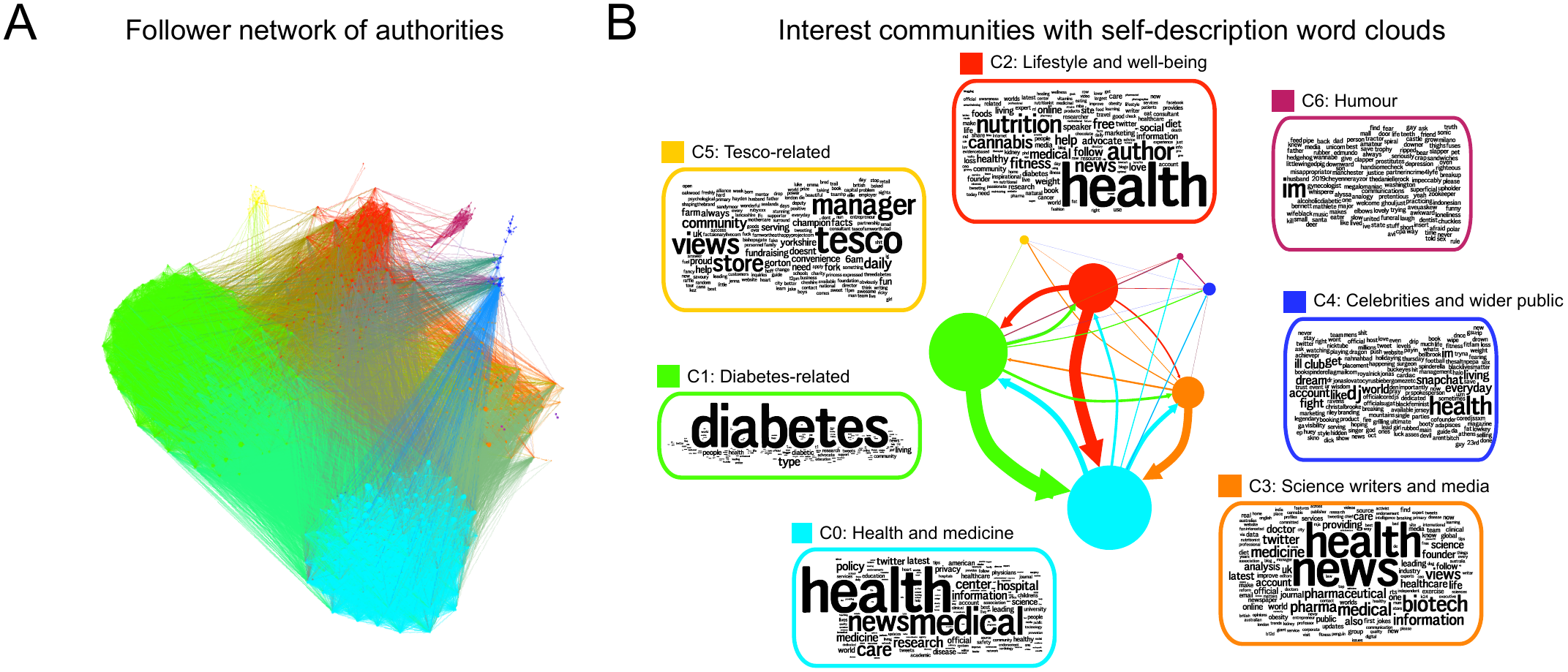}}
  \caption{{\bf A}: The Twitter follower network of the top
    authorities.  The nodes correspond to the 1000 users with the
    highest aggregate authority score and the follower network was
    obtained by mining Twitter in September 2015. The users are
    coloured according to their interest community, as obtained
    through the analysis of the follower network using Markov
    Stability. We find 7 main communities (there are an additional six
    very small communities not shown).  {\bf B}: The authority
    follower network coarse-grained by the obtained communities. Word
    clouds of the Twitter biographies of the users in each interest
    community were computed \textit{a posteriori} to help establish
    the thematic content of each of the seven groups, as summarised in
    the description.  The directed arrows reflect the direction of
    interest between groups, and the width of the arrows is
    proportional to the number of connections between the
    groups.}
  \label{fig:auth_communities}
\end{figure*}

This directed network has a large weakly connected component of 880
nodes (see Fig.~\ref{fig:auth_communities})---the rest of the nodes
are either isolated (a handful are connected to only one node) or have
since been deleted or blocked. We work below with this network of 880 
Twitter users and their follower relationships. 

It is important to note that, unlike the retweet networks studied up
to this point, the follower network is not `conversation-centric'. In
a retweet network, the existence of a connection is the result of one
user retweeting a message containing the term `diabetes' at least
once. In contrast, in the follower network a directed edge indicates
an interest by the `follower' to receive information from the
`followed' on a more permanent basis. This declaration of interest may
reflect more general information beyond the specific retweets about
diabetes.

As discussed above, the top 1000 authorities form a heterogeneous
group of users, including, among others: public health institutions
and foundations; diabetes advocates, researchers, activists, and
patients; hospitals and medical schools; academic and mainstream
publishers, media, and personalities; companies.  The most central
nodes in this follower network according to pagerank~\cite{Page1999,
  Gleich2015} (a proxy for importance) are users with a broad reach
and not diabetes-specific:
\begin{enumerate}
\item Centers for Disease Control \& Prevention (CDC), @CDCgov (ranked 65 as
  an authority in the previous section).
\item National Institutes of Health, @NIH (authority rank 111)
\item New York Times-Health, @NYTHealth (authority rank 150)
\item World Health Organization, @WHO (authority rank 88)
\item CDC flu updates, @CDCFlu (authority rank 87)
\item Recall information from the US Food and Drug Administration
  (FDA), @FDArecalls (authority rank 651)
\item News about medical research and health advice from the New York
  Times, @nytimeswell (authority rank 412)
\item National Public Radio-Health, @NPRHealth (authority rank 336)
\item US Office on Women's Health, @womenshealth (authority rank 93)
\item New England Journal of Medicine, @NEJM (authority rank 67)
\end{enumerate}
The high centrality of accounts relating to health research and information
(@NIH, @NEJM, @CDCgov, @WHO, @CDCFlu, @womenshealth), health-specific
news media (@NYTHealth, @nytimeswell, @NPRHealth), health advocacy
(@WHO) and commercial activities relating to foods and medicines
(@FDArecalls) indicates that health in general is a key concern to
users in this network.  Diabetes is one health concern, but it is not
isolated from others.

Beyond a ranking of centrality, it is important to extract information
about relevant communities in this follower network, so as to reveal
information about who the authorities are and about their interests.
A community in a network is usually defined as a highly cohesive group
of nodes, with above-expected connections within the
group~\cite{Porter2009a}.  In directed networks such as the follower
network, we define communities in terms of flows of information, i.e.
as groups of nodes in which interest or information is retained and
circulated. To extract such interest communities, we use the Markov
Stability framework~\cite{Delvenne2010,Delvenne2013}, a dynamical
framework especially well-suited to extract directed
communities~\cite{Lambiotte2014, Beguerisse2014}.  In this follower
network, we found a robust partition into seven main interest
communities (Fig.~\ref{fig:auth_communities}B) and six additional
communities with three nodes or fewer. Each of the seven larger
communities contains a distinct group of users characterised by domain
of activity, background, interests or employer (see Supplemental
Spreadsheet for the full list):
\begin{itemize}
\item[C0] \textbf{Health and medicine generalist accounts:} This
  community, the largest in the network (30\% of the nodes, light blue
  in Fig.~\ref{fig:auth_communities}), contains many accounts with
  high pagerank (including all the top 10 by pagerank listed
  above). Most members are related in general to health and
  medicine. For example, public health bodies, hospitals, medical
  schools, academic publishers, the health divisions of mainstream
  media, and commercial entities.

\item[C1] \textbf{Diabetes-related users:} This community (28\% of
  nodes, in green) contains the overwhelming majority of the
  diabetes-centred accounts discussed earlier, including associations,
  funders, advocates and patients, and companies. All of the top 10
  authority accounts in Fig.~\ref{fig:authorities} belong to this
  community.

\item[C2] \textbf{Lifestyle and well-being users:} This community
  contains accounts related to lifestyle and wellbeing advice,
  publications, natural and alternative remedies, detox, health and
  organic foods and products (17\% of the nodes, in red).

\item[C3] \textbf{Science writers and media:} This community contains
  science writers (e.g.,  @bengoldacre) and popular scientific
  publications, British media, and biotech-industry related accounts
  (11\% of the nodes, in orange).

\item[C4] \textbf{Celebrity and wider public:} This community contains
  celebrities, men's publications, and celebrity doctor @DrOz (5\% of
  nodes, in dark blue).

\item[C5] \textbf{Tesco-related:} In this community, most accounts are
  either owned by the UK retailer Tesco or by its employees (3\% of
  the nodes, in yellow).

\item[C6] \textbf{Humour:} This community contains comedians, parody,
  novelty accounts, and others (2\% of the nodes, in purple).
\end{itemize}

Through this succinct analysis of the follower network of authorities
we obtain a clearer global picture of: the general participants in the
diabetes-specific discussions analysed earlier; the groups of
interests that are present; the audiences involved in the debate; and
how they relate to each other.  For instance, as
Fig.~\ref{fig:auth_communities}B shows, the largest community (C0),
which contains generalist health and medicine agencies, is clearly
seen as a reference by all other communities, and especially by the
three other large communities: C1 (Diabetes), C2 (Life-style) and C3
(Science writers and media).  There are other clear asymmetries in the
follower-followed relationships: C2-lifestyle accounts strongly follow
C1-diabetes accounts (but this following is not reciprocated to the
same extent), whereas C2 users do not strongly follow C3-science
writers/media accounts (which largely ignore C2).  The connection
between C1-diabetes and C3-science writers/media is not strong in
either direction.  The presence of the UK retailer Tesco as a distinct
community (C5) in this network is also noteworthy; Tesco is a
high-profile supporter of the charity Diabetes UK (@DiabetesUK, in
C1)~\footnote{\url{ https://www.diabetes.org.uk/tesco/} Accessed on
  October 8 2015.}. As we have previously noted, humour is a prominent
feature in our datasets, and it makes another significant appearance
here. Community C6, which contains many comedy and parody
accounts. Parody and comedy communities have also been observed in
other follower networks obtained from different topical
issues~\cite{Beguerisse2014}.  To provide a visual interpretation of
the communities found, Figure~\ref{fig:auth_communities}B also
contains word-clouds constructed from the Twitter biographies of the
members in each community.
The word clouds are aligned with our descriptions of the interest
communities. In particular the members of communities C0, C1, C2, C3
and C5 use consistently-themed language to describe themselves. In
contrast, the members of the celebrity and humour communities (C4 and
C6) are heterogeneous in their self-descriptions, yet clearly
identifiable through examination of their usernames.  As noted
in~\cite{Beguerisse2014}, the word-clouds obtained from the
biographies can be thought of as an independent annotation or
`self-description' of the communities.  It is important to remark that
the biographies were not used in the analysis of the network.

\section{Discussion}

Twitter is a source of information and interaction for a growing
section of the world's population~\cite{Mitchell2015}.  Hence,
understanding online conversations around health issues on Twitter
(and social media platforms more broadly) is important, especially
since such platforms are frequently considered as possible tools in
public health outreach and health promotion
initiatives~\cite{Shiffman2012, Radmanesh2014, You2014, PHE2015,
  PHE2015a}.  Such initiatives are designed based on extended
assumptions about the Twitter health landscape, the messages which
dominate it, and the interactions of users within it. However,
detailed data-driven research can help ascertain the validity of such
assumptions and enrich our understanding of the online landscape and
its implications for population health.

In our study, references to diabetes on Twitter fall into four broad
thematic groups: health information, news, social exchanges and
commercial messages (Table~\ref{tab:topical_tweets}). While these
groups define a body of consistent themes, specific messages are
largely irregular and variable over time, both in their language
and content. 

In contrast, there exists an additional group of highly
consistent messages that are propagated through popular culture and
humour, including jokes, song lyrics or viral `facts'. Such messages
are consistent in content and style over time as they continue to be
posted and shared. Humorous tweets, banter, jokes and social
engagement (both supportive and stigmatising) are common in our
data. Tweets with such content have a different tone and vocabulary
than tweets containing authoritative posts, formal health messages or
news headlines.  This observation is in line with previous reports,
which have found that the most retweeted tweets are emotionally
evocative (either humorous or evoking anger)~\cite{Alvarez2015,
  Amor2015, So2015}. Moreover, we also find that these tweets appear
on Twitter more persistently over time than other tweets in the sample
(Fig.~\ref{fig:num_tweets}). From a health promotion perspective, this
highlights a need to consider not only the consistency of slogans or
lines in the messages, but also their style and sentiment. The
importance of sentiment in promoting dietary choices has also been
discussed in Refs.~\cite{McLennan2014, McLennan2015a, Ulijaszek2016}.

The abundance of jokes and sexual innuendo about foods and substances
that contribute to diabetes indicate at least a basic understanding of
diabetes, some of its causes, and its connection to blood sugar.  The
embedding of such fundamental understandings about diabetes in online
social media may be the result of health efforts in nutrition over the
past decades. This observation is at odd with assumptions that more
health education is required to help people to understand the sorts of
foods which might contribute to the development of
diabetes. Furthermore, such use of humour may imply a sense of
powerlessness to make `healthy' choices as users seemingly mock health
advice when faced with realities of the food and drink products they
encounter on a daily basis~\cite{McLennan2016}.

When it comes to the `who' of diabetes on Twitter, our analysis shows
a clear separation between the relatively few accounts that produce
the most engaging content (authorities), and the broader audiences
that disseminate and respond to it (hubs), as seen in
Fig.~\ref{fig:2D_hub_auth}). Only about 10\% of the accounts in our
dataset (approximately 120,000) produce tweets that evoke some
(heterogeneous) response (Fig.~\ref{fig:centralities-cdf}). Among the
top authority accounts, we find a mix of stockmarket-listed firms;
public, civil and grassroots organisations; and individuals who have
experience as diabetes patients and care-givers. The engagement levels
of all authorities are sustained throughout the observation period,
and it is difficult to clearly discern between these different groups
based on the content, style or theme of their tweets alone. These
observations reflect the fact that the health landscape today no
longer consists of government and citizens
alone~\cite{McLennan2015a}. Other entities (such as commercial agents,
individual bloggers and automated accounts) also exist and exert
influence on the health promotion landscape, and on the conversations
taking place around such themes. Their motives are sometimes
challenging to discern, and sponsorship arrangements can be difficult
to identify or regulate. Similar findings have been noted with respect
to Facebook, another social media platform used for health care and
communication~\cite{Greene2011}. Such entities, which are assumed by
health authorities to be extrinsic to the healthcare arena, need to be
acknowledged and considered in successful strategies for health
promotion.

Our results show that the role of hubs in disseminating information is
less persistent over time than the role played by authorities in
generating information.  Hubs are generally far less transparent about
their identity. Some accounts, for example, claim to have medical
expertise, but these claims cannot always be verified. As a result, it
is difficult for Twitter users to be `informed citizens' and discern
legitimate from misleading or discredited information, or a bona-fide
health expert from a social-media expert, or a marketer with business
motivations from a marketer with public health ones. Establishing the
credibility of an account by the number of messages, followers or even
through the published user profile, for example, can therefore be
misleading. This poses challenges to the use of Twitter as a health
promotion platform, and requires the use of sophisticated techniques
(e.g., collaboration of health professionals with practitioners of
network and data science). Further research and engagement with `real'
people, and not simply their virtual online personae which represent
them, is necessary to elucidate the expertise and intentions of users
generating some of the key diabetes-related content on Twitter.

This research has implications for health policy makers and health
promotion practitioners. We show that key authorities in relation to
diabetes on Twitter are not simply those with accredited and
formally-recognised health expertise, but also bloggers (men and
women), patients, celebrities, advocacy organisations,
stockmarket-listed firms, news media and automated accounts. There are
diverse stakeholders that use Twitter; even when limited to the
English language, these accounts originate from a range of countries
(notably, several Nigerian accounts have prominent roles as both hubs
and authorities).  It may be that establishing long-term collaborations
with the most influential users, as well as enhancing the connections
between different communities of users, may be more effective health
promotion strategies than running short-term government-led campaigns
and slogans aimed at informing people who have (or who are at risk of
developing) diabetes. In addition, it is important that policy makers
understand what social media users already know about health before
new initiatives are introduced: simply telling users what they already
appear to know, for example, about which foods contribute to diabetes,
may hinder health agencies' impact and credibility.

Our findings also have implications for
programme evaluation, where `number of tweets/followers' is commonly
assumed to equate to `impact'.  Other metrics that leverage the
structure of the network of interactions to consider relative impact,
or a sustained high-centrality presence over time (e.g., via centrality
metrics such as hub-authority scores) may more closely reflect the
reality of Twitter use for health purposes.

The health information landscape on Twitter is complex, and it cannot
be assumed that people can easily discern `good' and `bad'
information. Our observations echo previous reports that there is more
information available to consumers than they have the capacity to
process and understand~\cite{Berg2012, Gallotti2016}.  In this
context, public health approaches and messages that simply aim to
`inform' the public might be insufficient in themselves, or even be
counterproductive as they make a complicated cacophony of messages
even busier.  This is particularly relevant for health policy
makers. For instance, information that is disseminated by bloggers,
stockmarket-listed firms or automated accounts may be in line with
broad health recommendations (and indeed may provide a valuable
service to users), but without clear distinction from `legitimate'
health advice, such information might equally push particular aspects
of a commercial agenda that could lead to harm or greater health costs
in future. In this case, public health agencies may have to develop
novel approaches to ensure that the electronic health information
landscape is one that promotes healthy citizens and not only sweet
profits.

\section*{DECLARATIONS}

\subsection*{Contributions}
MBD and AM researched literature and conceived the study, carried out
quantitative (MBD) and qualitative (AM) data analysis, and prepared
the first draft of the manuscript. GGH and MBD collected the data.
MBD, AM, MB, SU wrote the paper.  All authors reviewed and edited the
manuscript and approved the final version of the manuscript.

\subsection*{Acknowledgements}

We thank E. Gardu\~no for useful advice and discussions on how best to
gather and process the data. We thank H. Harrington and S. Yaliraki
for fruitful discussions and comments on the manuscript.

\subsection*{Funding}
MBD acknowledges support from the James S. McDonnell Foundation
Postdoctoral Program in Complexity~Science/Complex~Systems Fellowship
Award (\#220020349-CS/PD Fellow).  MB acknowledges funding from the
EPSRC through grants EP/I017267/1 and
EP/N014529/1. \\

\subsection*{Data statement}
The IDs of the tweets used in this research and the Supplemental
Spreadsheet can be downloaded from the ReShare UK Data Service
repository~\cite{DiabetesTweets}.

\subsection*{Conflicting interests}
GGH is an employee of Sinnia, a data analytics company. The remaining
authors declare no conflicting interests.

\bibliographystyle{vancouver}

\appendix
\section{Additional tables with illustrative tweets for each thematic group}

\renewcommand{\arraystretch}{1.1}
\begin{table*}[tp]
  \smaller
  \begin{tabular}{|p{0.15\textwidth}|p{0.85\textwidth}|}
  \hline
  \multirow{2}{*}{\multititle{Public health messages}}
  &  Experts recommend universal diabetes testing for
  pregnant women at first prenatal visit http://t.co/6nLAUytBCX \\
  & FDA warns of massive diabetes test strip recall Nova
  Max strips http://t.co/dxtv4YLMIR via @nbcnewshealth \\
  \hline
  \multirow{6}{*}{\multititle{Links to articles, blogs and studies
      about risks, treatment and cure }}
  &  One Overlooked Trace
  Mineral Could Wipe Out Diabetes: { http://t.co/lD1GgBy6ry} \\
  & : Insulin Pumps vs. Insulin Injections for Type 1
  Children { http://t.co/d2Np519zAN} \#diabetes \\
  & Effect of imidapril versus ramipril on urinary
  albumin excretion in hypertensive patients with type 2
  diabetes... { http://t.co/wLZfIsiElp} \\
  &  Reverse Your Diabetes Today: Learn a little-known
  but 100\% proven way to erase your pre-diabetes and type 2
  diabetes. { http://t.co/TUDwKT3R8F} \\
  & Studies show chlorella could improve insulin
  sensitivity in type 2 diabetes patients { http://t.co/J4GxrbLF32}
  via @HealthRanger \\
  & Flu shot extra important if you have diabetes \/\/
  { http://t.co/59Qx57TiI4}\\
  \hline
  \multirow{4}{*}{\multititle{Population health and fears}}
  & Health chief fears diabetes to soar across
  Bradford district (From Bradford Telegraph and Argus)
  http://t.co/g5A1yZTbsJ \\
  &  Fears Rockhampton facing diabetes epidemic: Diabetes
  Queensland says more than 10 per cent of the population
  i... http://t.co/bQhnR4x4Rk\\
  & Obesity and diabetes pose a serious threat to the
  long-term health of young people in the United
  States. http://t.co/biz4U2gWup\\
  & Diabetes is a disease that can strike when you
  don't take care of your body. Check out these eye-opening
  statistics. http://t.co/zwpfTPgbtu\\
  \hline
  \multirow{5}{*}{\multititle{Publicity about outreach and awareness
      events and activities}}
  & Many @Enterasys
  employees riding in @AmDiabetesAssn Tour de Cure are too familiar w/
  diabetes. Here's why they ride { http://t.co/qURu4XP2xy} \\
  &  Donate to @Brenda\_Novak's auction for the cure for
  diabetes. I'm giving away 2 tix to my blog tour course \#win - {
    http://t.co/sR65gINrx6} \\
  & Join us as we fight
  to \#StopDiabetes. Sign our petition to urge Congress to invest in
  \#diabetes prevention \& a cure: { http://t.co/6ztkHxnNy1} \\
  &  Support World Diabetes Day 2012, add a \#twibbon to your
  avatar now! - { http://t.co/7mLvcvY1Hx} \\
  &  Merck Animal Health Launches Global Awareness Campaign to Support
  Pet Diabetes Month\texttrademark~{ http://t.co/BDr77CoCg5} \\
  \hline
  \multirow{4}{*}{\multititle{Advice
      about diabetes management and diagnosis}} 
  & Open
  Question: Diabetes question i want to know if i have diabeties? i
  was in eye hospital today and they have... {
    http://t.co/2BNfwNZL2f}\\
  & Open Question:
  Whart are some good ways to deal with pre-diabetes? 9I have some too
  but are open to yous)? { http://t.co/y6wT3BXSaX} \\
  & 5 things \#caregivers need to know about
  \#diabetes. { http://t.co/oncoPhK0QH} via @sallyabrahms \\
  & Open Question: I think I may have diabetes?  {
    http://t.co/UGnLZaZ3Gf} \\
  \hline
  \multirow{7}{*}{\multititle{Lifestyle, diet and cookery tips, news and links}}
  & 8 Tips for Eating Out With Diabetes - Type 2
  Diabetes Center-Everyday Health { http://t.co/u5nIZ4cg5E}
  \#diabetes \#health \#diettips \\
  & Control Your
  Diabetes: Diet Tips Check out this post Control Your Diabetes: Diet
  Tips!.. { http://t.co/nES3pfIwg} \\
  & Purina
  veterinary diets dm { http://t.co/cJ7bVFrbz8} \#diabetes
  management feline formula \\
  &  Zucchini Escarole Soup:
  From Diabetes Cooking for Everyone, by Carol Gelles. Exchanges: 1/4
  bread, 1 vegetable,... { http://t.co/IiH1Da9d4u} \\
  & Micronutrient Enriched Wheat Steamed Bread is
  Beneficial for Diabetes Patients { http://t.co/SHthuJliHh}
  \#health \#cancer \\
  & Tips For Living A Life
  With Diabetes: TIP!  Almonds are a great way to get some additional
  protein into your di... { http://t.co/XPkJm6jAJ0} \\
  & To manage your diabetes, Weird Science recommends
  the munchies { http://t.co/QybNo8qfTd} \\
  \hline
  \multirow{4}{*}{\multititle{Life stories and experiences (some for
      marketing purposes)}} 
  & Tiny cells lift Type 1 diabetes
  hope: Michael Schofield gets to meet the mother of the man whose
  death gave hi... { http://t.co/ODMOzctOWI} \\
  & Debby M Shared this yesterday and its outstanding!!!
  April 23rd 2013, I was
  diagnosed with diabetes type 2. My... { http://t.co/iv4blIR5Nj} \\
  & The Day I was Diagnosed with Juvenile Diabetes: I
  was diagnosed on July 21st 1999 when I was eight years old. ... {
    http://t.co/U1GPaX812b} \\
  &  \#50ThingsAboutMe  38) I have diabetes since
  I was 9 years old\\
  \hline
  \multirow{6}{*}{\multititle{Dangers of sugar, sugar replacements
      and/or soda}} 
  &  Drinking one can of soda a day
  increases your chances of getting type 2 diabetes by 22\%! Details
  -- { http://t.co/BzOmFKhhbw} \\
  &  Diabetes warning over soft sugary drinks: Dr Tim
  Dalton, chair of the Wigan Borough Clinical Commissioning
  Gro... { http://t.co/uaLWTZaHDI} \\
  & A soda a day keeps the doctor in pay: soft drinks and
  diabetes: Recent research linking soft drinks to type 2 ... {
    http://t.co/QhB31VRES2} \\
  & If Mountain Dew: Baja Blast was sold in the stores,
  diabetes would spread like wildfire, y'all. \\
  & Thanks for putting a damper on my \#1 pregnancy
  craving. Coke and Pepsi Face Diabetes Backlash {
    http://t.co/MI8it8xphD} via @adweek \\
  & Fighting flab? Think before u reach out 4 sugar
  substitute \#Sugar \#Sucralose \#Diabetes {
    http://t.co/cfeYURK7r3} \\
  \hline
\end{tabular}

  \caption{Specific examples of health information tweets.}
  \label{tab:healthtweets}
\end{table*}

\renewcommand{\arraystretch}{1.1}
\begin{table*}[tp]
  \smaller
  \begin{tabular}{|p{0.15\textwidth}|p{0.85\textwidth}|}
  \hline
  \multirow{4}{*}{\multititle{Headline links to particular
      `breakthrough' studies or technologies}}
  & Immune protein could stop diabetes in its
  tracks, discovery suggests: Researchers have identified an immune
  pr... http://t.co/rd8Jx2avnn\\
  & \$FPMI - Advanced Imaging Studies May Enhance Diabetes
  Management http://t.co/Ow3QX0yuYm\\
  & Fish Oil Pills Might Cut Diabetes Risk,
  Researchers Say http://t.co/fvsIMXFKGR\\
  & Great article from @jdwilson2 on islet cell transplant
  testing offering hope for T1 \#diabetes http://t.co/qn9XYwIxC5
  @cnnhealth\\
  \hline
  \multirow{3}{*}{\multititle{Celebrity news}}
  &  Another reason to dislike the Pats | NFL Patriots
  release Kyle Love after diabetes diagnosis: http://t.co/lAV114TP06
  via @wtcommunities\\
  &  Sherri Shepherd Talks About Her Fight With
  Diabetes! [Video]: video platformvideo managementvideo
  solutionsvid... http://t.co/6FiJMcUal3\\
  & Wow Trump just gave Lil John \$100K toward his
  diabetes charity \& his mom just passed away they all cheered
  endlessly. \#apprenticefinale\\
  \hline
  \multirow{4}{*}{\multititle{General news articles about diabetic
      people or pets}}
  &  Thief steals family car with daughters \#diabetes
  medicine insidehttp://youtu.be/juPZtMmzL2s via @youtube @tmz
  http://t.co/agDquGWmUL\\
  & Diabetes service dog saves the day and night
  for Fort Worth teen http://t.co/LQ4NRm9PLm\\
  & Cop beats up girl for diabetes, exposes racist
  government schools: http://t.co/5AYnBxPEsI via @youtube \\
  & Why Jasper TX Service Trains Quality Service Dogs For
  Those With Diabetes And Autism http://t.co/xFzI5hdRCP\\
  \hline
  \multirow{6}{*}{\multititle{News relating to the pharmaceutical
      industry and the economy}}
  & Glooko's New Diabetes Management System FDA Cleared:
  Glooko (Palo Alto, CA) received FDA clearance for its
  lat... http://t.co/j1KffOmigz\\
  &  \$CXM Plays Two Vital Roles in the Fight Against
  Diabetes http://t.co/f0zVaXMsv9 | \#diabetes \#stocks
  http://t.co/QK5Zw5tGQt\\
  & Woman with Type 2 diabetes sees premiums plummet from
  \$500 to \$1 under Obamacare | http://t.co/CUurbdnwqR \\
  &  News: Cytori Licenses AsiaPacific Cardiovascular
  Renal \& Diabetes Markets to Lorem Vascular for up to \$531 Million
  http://t.co/6SNySsC7qX\\
  &  AstraZeneca could buy Bristol stake in diabetes JV:
  analyst: LONDON (Reuters) - AstraZeneca may seek to
  increa... http://t.co/rcZQPIIZxo \\
  & Salix to Buy Santarus for \$2.6 Billion to Gain
  (type 2) \#Diabetes Drug \mbox{http://t.co/wpjNIxKGLf \#} \\
  \hline
\end{tabular}

  \caption{Examples of news tweets.}
  \label{tab:newstweets}
\end{table*}

\renewcommand{\arraystretch}{1.1}
\begin{table*}[tp]
  \smaller
  \begin{tabular}{|p{0.15\textwidth}|p{0.85\textwidth}|}
  \hline
  \multirow{4}{*}{\multititle{Users joking about how what they
      have eaten is likely to give them diabetes}}
  &  cake cake cake cake cake cake cake cake cake cake cake
  cake cake cake cake cake cake cake cake cake cake cake cake cake
  cake cake diabetes \\
  &  2 bowls of yogurt, a bowl of oreos, hersheys,
  chips, cheese and a shitload of mints. My diet consists of
  diabetes. \\
  &  Bother! Burger King has arrived. Hello obesity,
  diabetes, poor nutrition. McD's is bad enough. Grumble Grumble
  ????\\
  &  To us: candy,chocolate,nasi lemak,fries To parents:
  diabetes,tooth decay,heart disease,high blood pressure.\\
  \hline
  \multirow{2}{*}{\multititle{Chatter and social interchanges}}
  &  I JUST TOLD MY DAD A BAD DIABETES JOKE AND HE'S A
  DIABETIC OMFG IM SUCH A BAD PERSON \\
  &  My mom said I probably have diabetes.. aaa I didn't
  know she went to school to be a doctor lol\\
  \hline
  \multirow{7}{*}{\multititle{Everyday experiences of diabetes}}
  &  Screw this diabetes business today. Just wanna
  sleep.. \#HeadAche \#FeelHungover\\
  &  My \#husband going in for \#surgery this morning
  for \#amputation of his toe and partial foot due to \#diabetes
  \#diabetesawareness \#fckudiabetes\\
  &  Watching belly dancers on youtube \& wishing I
  never stopped going-No time \& damn diabetes took over my
  life. \#killjoy \#fcbd \#hipsofglory\\
  &  Breakfast gone wild? Pump set gone wild? Hellooo
  Monday. WTF diabetes. http://t.co/uK3xTuJuMD\\
  &  Me and diabetes aren't getting along
  today. \#ifeellikepoop \#badmood \#diabeticproblems ????
  http://t.co/3OwXSe2keO\\
  &  Ever have a cat chew on your insulin pump
  tubing? \#diabetes \#tubing \#insulinpump\\
  &  Is it too much to ask that media clearly differentiate
  between type 1 and type 2 diabetes? Lap band surgery will not cure
  T1 diabetes \#fedup \\
  \hline
  \multirow{7}{*}{\multititle{Stigmatising comments}}
  &  ``I'm working on something'' ``The only thing
  you're working on is diabetes you fat fuck'' \#Projectx is jokes
  :L\\
  &  Why do you think your beautiful
  you look like a fucking frog with diabetesOMFG ahahahahahahahah\\
  &  To the cunt in work that smeared shit on the floor in
  the toilets. I hope you get type 2 diabetes and aids.\\
  & I hope y'all fucking future kids get fat and get
  diabetes talking shit bout Ms Dumas\\
  &  I seriously miss my doctor in touness, this one told
  me today that he thinks diabetes is gonna be the cause of my
  death. How rude):\\
  &  `A nation of porkers': Diabetes expert complains on
  national radio that we're eating ourselves into an early grave\\
  &  Excuse me sir..Just leave the gym with ur
  diabetes-lookin, wheezin self.. Puffing ur chest doesnt hide ur huge
  beer belly ??? \#idontplayatgym \\
  \hline
  \multirow{6}{*}{\multititle{Sexual innuendo and humour relating to
      sweetness and diabetes}}
  &  'Niall's so sweet, i'd bet you'd get diabetes by
  swallowing his cum' OMFG HAHAAHAHAAH I LOVE THIS\\
  &  HARRY YOU LJTTLE CUTIE PATOOTIE I THINK I JUST RECIEVED
  DIABETES FROM LOOKIGN AT THIS SUGAR PIE HOINEY BUCNH
  http://t.co/nJZqYUfOMW\\
  &  I'd lick your sweet pussy till I get diabetes.\\
  &  Do me baby. Uh oh, did you say DEW ME?! *Mountain
  Dew gushes from his penis and gives her sexually transmitted
  diabetes*\\
  &  woah woah stop giving girls diabetes Ahh EYE CANDYMAN
  -- EH I don't even know what's going on now! What eye candy...\\
  &  \#ABCReports The Blacker The
  Berry,The Sweeter the Juice Is Wesley Snipes own sweat giving him
  Diabetes???More @ 10pmlol \\
  \hline
  \multirow{2}{*}{\multititle{Jokes, sarcasm and humorous tweets}}
  &  Just saw a commercial for diabetes
  medicine. Side effects: Low Blood Sugar. Let that sink in.\\
  &  I didn't even know what the gd pancreas did
  until Violet was diagnosed with Type 1 Diabetes. Fuck you, you
  shitty non-working pancreas.\\
  &  A marwadi opens a Sweets Shop ...  Puts an
  Advertisement : Helper Needed ...  Qualification : Should have
  Diabetes\\
  &  Pray for the rain forest. Pray for that gay NBA
  player. Pray for the o-zone layer. Pray for stray animals. Pray for
  diabetes\\
  &   The coca cola Christmas advert,
  because nothing says Christmas quite like diabetes and
  capitalism. LOL\\
  \hline
\end{tabular}

  \caption{Social interaction and humorous tweets.}
  \label{tab:SocialInteraction_Humour}
\end{table*}

\renewcommand{\arraystretch}{1.1}
\begin{table*}[tp]
  \smaller
  \begin{tabular}{|p{0.15\textwidth}|p{0.85\textwidth}|}
  \hline
  \multirow{4}{*}{\multititle{Advertisements for jobs in the
      pharmaceutical and care industries}}
  & Sioux City Jobs: Sioux City, IA Diabetes Sales
  Specialist at Inventiv Health (Sioux City, IA)
  http://t.co/9vDO6xjUKS \#Jobs \#SiouxCityJobs\\ 
  &  \#jobs4u \#jobs \#ABQ Pharmaceutical Representative -
  Diabetes Products - Albuquerque, NM http://t.co/ghNCYrMbHo
  \#albuquerque \#NM\\  
  &  \#jobs,\#ukjobs Clinical Nurse Specialist Diabetes
  http://t.co/QtnCwYK6WR \#jobs4u\\ 
  &  Start a new \#career at American Diabetes Association
  in Rocky Hill, CT.  Associate Director -
  Fundr... http://t.co/Wi3tM70wMd \\ 
  \hline
  \multirow{3}{*}{\multititle{Marketing for a specific product, app,
      treatment, event or service}}
  &  Caffeine stimulates elevated of Cortisol =
  arthritis, obesity, diabetes, and depression. Try healthy coffee: -
  http://t.co/4paSPZ0mrc\\ 
  &  Acid Reflux, Arthritis, Diabetes, Enlarged Prostate,
  Overweight! http://t.co/aCXiugNQsh\\ 
  &  FREE Kindle eBook: Apple Cider Vinegar Natural Cures
  for Diabetes, Cancer, and MORE!... http://t.co/gGNGXwVfqx\\
  \hline
  \multirow{5}{*}{\multititle{Buy diabetes drugs, diets or treatment
      products online}}
  &  http://t.co/wJ6z0hd1BO Buy Diabetes online if
  Cheap Diabetes no prescription, Order Diabetes without
  prescription\\ 
  &  aggressive diabetes actos http://t.co/biK2GYcOQI
  \#buy \#cheap \#pills \#online \#pharmacy \#drugs \#generic\\ 
  &  \#aldactone: Co-trimoxazole: Buy Generic Bactrim -
  Aldactone for diabetes at http://t.co/0qM9gUUa69\\ 
  &  No Script Slimex For Cheap No Prescriptions
  Needed For Slimex Cause Diabetes http://t.co/6uhu2RmMZh\\ 
  &  American-Diabetes-Wholesale : \$12 Off Order of
  \$100 or More! Code: ADW12100 http://t.co/e5K20ptIhH\\
  \hline
\end{tabular}

  \caption{Commercial tweets.}
  \label{tab:Commercial}
\end{table*}

\renewcommand{\arraystretch}{1.1}
\begin{table*}[tp]
  \smaller
  \begin{tabular}{|p{0.15\textwidth}|p{0.85\textwidth}|}
  \hline
  \multirow{2}{*}{\multititle{Lyrics of `All the time', by
      Jeremih~\cite{Jeremih2013}}}
  & If its sweet then ima eat it, til I get sugar
  diabetes. Ima blood n she anemic. We perfect.\\
  & And If It's Sweet Then Ima Eat It Till I Get
  Sugar Diabetes Ima Blood And She A Nemick!!!!\\
  &  And if it''s sweet then imma smoke it till I get sugar
  diabetes - wayne\\
  & Pusssssy for breakfast. Yo' pussy betta. Ima eat
  till i get sugar diabetes. \#AllTheTime \#Jermiah \\
  \hline
  \multirow{3}{*}{\multititle{Lyrics of `Mind of a maniac', by
      Boosie Badazz~\cite{Boosie2010}}}
  &  Use to hold my head down.. not no more diabetes n
  my body police kickin in my door.. but I'm still
  happyyyy... \#boosie\\
  &  USED TO HOLD MY HEAD DOWN NOT NOMO DIABETES IN MY BODY
  POLICE KICKING IN MY DOE!!!BUT IM STILL HAPPY!!!!!!\\
  &  diabetes in my body police kickin in my door, but
  im still happy.\\
  \hline
  \multirow{3}{*}{\multititle{Alcohol reduces diabetes risk}}
  & ``One Alcoholic drink a day can reduce
  your risk of type 2 diabetes by up to 30 percent.'' See I'm not an
  alcy I'm keepin helthy\\
  & ``One Alcoholic
  drink a day can reduce your risk of type 2 diabetes by up to 30
  percent.'' Let's drink more\\
  & ``One Alcoholic drink a day can
  reduce your risk of type 2 diabetes by up to 30
  percent.''  \#tequila\\
  \hline
  \multirow{3}{*}{\multititle{Viral fact about detecting diabetes by tasting
      urine}}
  & ``Doctors used to taste urine to
  determine if someone had diabetes because their urine would taste
  sweet.'' HAHAHAHA\\
  & ``Doctors used to taste urine to
  determine if someone had diabetes because their urine would taste
  sweet.'' \#nasty\\
  & Ammm ewwww ``Doctors used to taste
  urine to determine if someone had diabetes'' \\
   \hline
  \multirow{3}{*}{\multititle{The mathematics joke}}
  & Elementary math problems are weird. 'I had 10
  chocolate bars and ate 9. What do I have now?' Oh, I don't know,
  DIABETES MAYBE. :|\\
  &  Louis has 40 chocolate bars. He eats
  35. What does Louis have now? Diabetes. Louis has diabetes.\\
  &  Here's a question. Juan has 40 choco bars. He eats
  35. What does Juan hve now? Diabetes. Juan has
  diabetes. *Mathematics + Logic = Sarcasm\\
  \hline
\end{tabular}

  \caption{Recurrent tweets}
  \label{tab:Recurrent}
\end{table*}

\end{document}